\documentclass[twocolumn]{aastex63}

\received{December 14, 2021}
\revised{April 18, 2022}
\accepted{June 8, 2022}

\submitjournal{ApJ}

\shorttitle{Star Formation and Black Hole Accretion in LIRGs}
\shortauthors{Stone et al.}

\usepackage{enumitem}
\setlist{nosep}

\usepackage{amsmath}

\graphicspath{{./}{figures/}}

\begin{document}

\title{Measuring Star Formation and Black Hole Accretion Rates In Tandem using Mid-Infrared Spectra of Local Infrared-Luminous Galaxies}

\author[0000-0002-9720-3255]{Meredith A. Stone}
\affiliation{Department of Astronomy, University of Massachusetts, Amherst, MA 01003, USA}

\author[0000-0001-8592-2706]{Alexandra Pope}
\affiliation{Department of Astronomy, University of Massachusetts, Amherst, MA 01003, USA}

\author[0000-0002-6149-8178]{Jed McKinney}
\affiliation{Department of Astronomy, University of Massachusetts, Amherst, MA 01003, USA}

\author[0000-0003-3498-2973]{Lee Armus}
\affiliation{Spitzer Science Center, California Institute of Technology, Pasadena, CA 91125, USA}

\author[0000-0003-0699-6083]{Tanio D\'{i}az-Santos}
\affiliation{Institute of Astrophysics, Foundation for Research and Technology-Hellas (FORTH), Heraklion, GR-70013, Greece}

\author[0000-0003-4268-0393]{Hanae Inami}
\affiliation{Hiroshima Astrophysical Science Center, Hiroshima University, 1-3-1 Kagamiyama, Higashi-Hiroshima, Hiroshima 739-8526, Japan}

\author[0000-0002-5537-8110]{Allison Kirkpatrick}
\affiliation{Department of Physics and Astronomy, University of Kansas, Lawrence, KS 66045, USA}

\author[0000-0002-2596-8531]{Sabrina Stierwalt}
\affiliation{Physics Department, Occidental College, 1600 Campus Road, Los Angeles, CA 90041 USA}

\begin{abstract}

We present the results of a stacking analysis performed on \textit{Spitzer}/Infrared Spectrograph high-resolution mid-infrared spectra of luminous infrared galaxies (LIRGs) in the Great Observatories All-Sky LIRG Survey (GOALS). By binning on mid-infrared active galactic nucleus (AGN) fraction and stacking spectra, we detect bright emission lines [Ne~II] and [Ne~III], which trace star formation, and fainter emission lines [Ne~V] and [O~IV], which trace AGN activity, throughout the sample. 
We find the [Ne~II] luminosity is fairly constant across all AGN fraction bins, while the [O~IV] and [Ne~V] luminosities increase by over an order of magnitude. Our measured average line ratios, [Ne~V]/[Ne~II] and [O~IV]/[Ne~II], at low AGN fraction are similar to H~II galaxies while the line ratios at high AGN fraction are similar to LINERs and Seyferts. 
We decompose the [O~IV] luminosity into star-formation and AGN components by fitting the [O~IV] luminosity as a function of the [Ne~II] luminosity and the mid-infrared AGN fraction. The [O~IV] luminosity in LIRGs is dominated by star formation for mid-infrared AGN fractions $\lesssim0.3$. With the corrected [O~IV] luminosity, we calculate black hole accretion rates ranging from $10^{-5}$ M$_{\odot}$/yr at low AGN fractions to 0.2 M$_{\odot}$/yr at the highest AGN fractions. We find that using the [O~IV] luminosity, without correcting for star formation, can lead to an overestimate of the BHAR by up to a factor of 30 in starburst dominated LIRGs. Finally, we show the BHAR/SFR ratio increases by more than three orders of magnitude as a function of mid-infrared AGN fraction in LIRGs.
\end{abstract}

\section{Introduction} \label{sec:intro}

Understanding the interplay between star formation and supermassive black hole growth within galaxies is critically important to understanding how galaxies evolve over cosmic time. Both of these processes impact the gas in galaxies; by heating it or expelling it from the galaxy entirely, affecting the galaxy's future potential for growth.
While UV and optical studies can trace the unobscured star formation and active galactic nucleus (AGN) activity, the infrared (IR) is critical in dusty galaxies in order to detect the majority of the light from both processes that is absorbed by interstellar dust and re-emitted at longer wavelengths \cite[e.g.][]{Madau2014}. 

The mid-infrared (mid-IR) emission of galaxies hosts a number of spectral features sensitive to the ongoing star formation, AGN activity and conditions of the interstellar medium \citep[ISM, e.g.][]{Smith2007, Brandl2006, Lacy2020}. The atomic emission lines in the mid-IR [Ne~II] 12.8$\,\mu$m  and [Ne~V] 14.3$\,\mu$m are excited primarily by star formation and AGN, respectively \citep{Spinoglio1992,Ho2007,Veilleux2009,Tommasin2010,Dasyra2008}, and are thus good tracers of the star formation rate (SFR) and the black hole accretion rate (BHAR). The most promising BHAR tracer in the mid-IR is [Ne~V], due to its high (97.1 eV) ionization potential, above the limit of ionizing radiation produced by O and B stars. However, the mid-infrared [Ne~V] lines at 14.3 and 24.3$\,\mu$m are faint and often only detected in the strongest AGN sources \citep{Veilleux2009,Inami2013,Armus2006}.  
On the other hand, lines with intermediate ionization potentials such as [Ne~III] 15.5$\,\mu$m and [O~IV] 25.9$\,\mu$m may be excited by a combination of these processes, which complicates their use as SFR or BHAR tracers. [O~IV] can be used reliably as a tracer of AGN activity in galaxies with strong AGN such as those with high [O~IV]/[Ne~II] ratios \citep[e.g.][]{Tommasin2008,Goulding2010,DiamondStanic2012,AlonsoHerrero2012}, but the starburst component becomes increasingly important in galaxies with weaker emission from AGN relative to that from star formation \citep{Armus2006,PereiraSantaella2010}. In order to use [O~IV] to quantitatively measure AGN properties in composite systems, its starburst component must be removed. A correction using e.g. [Ne~II] to estimate the starburst component must be calibrated across a large sample of galaxies representing the full range of AGN strengths. However, most previous studies have not detected all three lines, [Ne~II], [Ne~V] and [O~IV], in individual galaxies spanning the full range of AGN contributions \citep[e.g.][]{Inami2013}. 

Our approach to overcome the challenge of faint or undetected spectral lines is to employ a spectral stacking analysis on the mid-infrared spectra of a well-defined sample of galaxies. Nearby luminous infrared galaxies (LIRGs, $10^{11} L_{\odot} \leq L_{\rm{IR}, 8-1000 \mu \mathrm{m}} \leq 10^{12} L_{\odot}$) and ultra-luminous infrared galaxies (ULIRGs, $L_{\rm{IR}} > 10^{12} L_{\odot}$) are an ideal population to study the coevolution of star formation and AGN accretion, because their mid-IR spectra show that they host both strong star formation and significant AGN activity \citep{Genzel1998,Petric2011}.

The Great Observatories All-Sky LIRG Survey \citep[GOALS,][]{Armus2009} is a complete subset of LIRGs and ULIRGs from the IRAS Revised Bright Galaxy Sample \citep[RGBS,][]{Sanders2003}. This sample has been extensively studied at a wide range of wavelengths, including mid-IR spectroscopy with the \textit{Spitzer Space Telescope} Infrared Spectrograph \citep{Petric2011,Stierwalt2013,Inami2013,DiazSantos2010,DiazSantos2011}. \cite{Petric2011} used multiple spectral diagnostics, including fine-structure lines, Polycyclic Aromatic Hydrocarbon (PAH) features, and the mid-infrared continuum to measure the mid-IR AGN fraction---the fraction of a galaxy's mid-IR emission ($\sim$8-40 $\mu$m) produced by its AGN---in the GOALS sample, finding that 18\% contain a strong AGN. However, in addition to these strong AGN, evidence suggests that a larger fraction of the sample contains sources with contributions from an AGN to their emission at lower levels. \cite{Stierwalt2013} utilized the equivalent width of the 6.2 $\mu$m Polycylic Aromatic Hydrocarbon (PAH) feature, which is inversely proportional to the AGN fraction \citep{Tommasin2008,Veilleux2009}, to demonstrate that approximately half of the GOALS sample either contains a strong AGN or is a ``composite" source with measurable contributions from both a starburst and a weaker AGN. Similarly, while the unambiguous but faint tracer of AGN, [Ne~V], is undetected in the majority of the sample \citep{Inami2013}, [O~IV] is detected in slightly more than half of the sample. Studies at other wavelengths, including X-rays \citep[e.g.][]{Iwasawa2011,Torres2018} and radio \citep{U2012,Vardoulaki2015} find similar results, that approximately half of the GOALS sample are either strong AGN or composite sources. In order to tease out these fainter AGN and quantify their BHARs, it is critical to detect AGN sensitive lines across the full sample and account for the contribution from star formation to lines like [O~IV]. 

Our goal is to simultaneously detect both the star formation and AGN mid-infrared lines in a large enough sample of LIRGs to investigate the relative SFR and AGN contributions and robustly estimate the BHARs. 
Given the range of IR luminosities, star formation rates, and AGN contributions in the GOALS sample, the undetected lines are likely just below the noise level in the spectra. 
In this paper, we stack the high resolution {\it Spitzer}/IRS spectra of the GOALS sample in bins of mid-IR AGN fraction with the aim of reducing the noise and detecting fainter lines such as [Ne~V]. 
We explore the evolution of [Ne~II], [Ne~III], [Ne~V], and [O~IV] across the full range of mid-IR AGN fraction, and decompose the [O~IV] line into contributions from star formation and the AGN.
Our sample and data are described in section \ref{sec:sample}, and our analysis in section \ref{sec:analysis}. We present our results in section \ref{sec:results}, and examine the evolution of star formation and black hole accretion rates in section \ref{sec:discussion}. We summarize and discuss future applications of this work in section \ref{sec:summary}.

\begin{figure*}[ht]
    \plotone{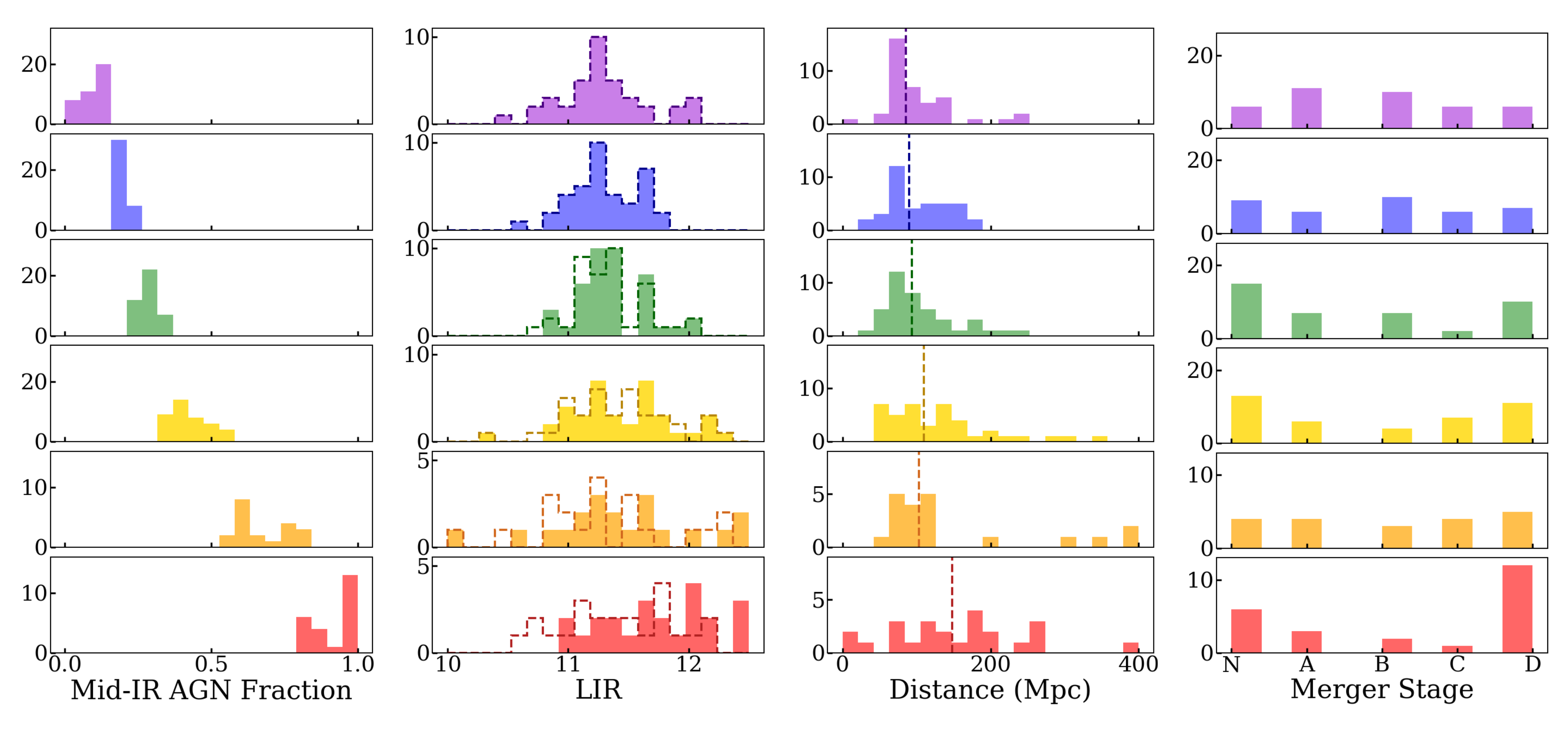}
    \caption{Distribution of mid-IR AGN fraction \citep{DiazSantos2017} from 6.2 $\mu$m PAH equivalent width, \citep{Stierwalt2013}, IR luminosity \citep{DiazSantos2017}, luminosity distance \citep{Armus2009}, and merger stage \citep{Stierwalt2013} for our six AGN fraction bins. We use different colors to denote the 6 bins in AGN fraction from purple, at the lowest AGN fractions, to red, at the highest AGN fractions. In panel 2, the total IR luminosity, L$_{\rm{IR}}$, is the solid distribution while just the IR luminosity from star formation, L$_{\rm{IR, SF}}$ is the dashed distribution. The vertical line in panel 3 shows the median distance for each bin. The merger stages from \cite{Stierwalt2013} are N (non-mergers), A (pre-merger), B (early-stage merger), C (mid-stage merger), and D (late-stage merger).
    \label{fig:bincharacteristics}}
\end{figure*}

\section{Sample and Data} \label{sec:sample}
Our sample consists of GOALS galaxies with high-resolution ($R=600$) {\it Spitzer}/IRS data in both short-high (SH) and long-high (LH) channels \citep{Inami2013}. We retrieved the 1D reduced spectra from the Infrared Science Archive \citep{GOALS}. These spectra were extracted using a full-slit extraction, and dedicated backgrounds were observed for subtraction: further details of the data reduction can be found in \cite{Inami2013}.

We excluded 10 GOALS galaxies with very noisy spectra. 
In addition, we require each galaxy to have a mid-IR AGN fraction calculated from the equivalent width (EQW) of the 6.2$\,\mu$m polycyclic aromatic hydrocarbon (PAH) line; these are tabulated in  \cite{DiazSantos2017} and \cite{Stierwalt2013}. We use this AGN fraction indicator because it measures the relative strength of the AGN in the mid-IR \citep[e.g.][]{Desai2007}, where the AGN contributes significantly, and because it does not depend on the mid-IR atomic lines that we aim to measure. We remove four galaxies whose PAH EQW-implied mid-IR AGN fractions were inconsistent with other AGN indicators, mainly the mid-infrared continuum slope, and thus added substantial noise to the stacks if included. Our final sample, after these 14 nuclei are removed, consists of 203 galaxy nuclei from the GOALS survey. The luminosity distances, total IR luminosities (8--1000$\,\mu$m, L$_{\rm{IR}}$) and merger stages for each galaxy are retrieved from \cite{Armus2009}, \cite{DiazSantos2017} and \cite{Stierwalt2013}. Since the AGN will contribute to L$_{\rm{IR}}$, we use the relation from \cite{Kirkpatrick2015} derived from mid-IR spectra of LIRGs at $0.3 \leq z \leq 2.8$ to convert the mid-IR AGN fraction to a bolometric AGN fraction, and use it to determine the fraction of L$_{\rm{IR}}$ from star formation: L$_{\rm{IR,SF}}$. Bolometric AGN fractions of the GOALS sample derived following the formalism of \cite{Veilleux2009} are tabulated in \cite{DiazSantos2017}. These agree well with the bolometric AGN fractions derived from \cite{Kirkpatrick2015} up to a mid-IR AGN fraction of $\sim0.9$ (which corresponds to a bolometric AGN fraction of $\sim0.5$), above which the \cite{DiazSantos2017} bolometric AGN fractions increase very steeply up to 1. As the \cite{DiazSantos2017} AGN fractions are based on templates of sources that more closely resemble Palomar-green (PG) quasars and the GOALS sample is IR-selected, we elect to use the \cite{Kirkpatrick2015} relation to derive bolometric AGN fractions and L$_{\mathrm{IR,SF}}$ but note that our general results and the trends in the data are not strongly affected by the small number of sources with the highest AGN fractions.

\section{Analysis} \label{sec:analysis}
In this work, we study the relationships between the luminosities of 12.8$\,\mu$m [Ne~II], 14.3$\,\mu$m [Ne~V], 15.5$\,\mu$m [Ne~III], and 25.9$\,\mu$m [O~IV] as a function of mid-IR AGN fraction. 

\subsection{AGN fraction bins}

For the analysis in this paper, we split our 203 galaxies into six bins of mid-IR AGN fraction. We choose to use the mid-IR AGN fraction derived from the equivalent width of the 6.2$\,\mu$m PAH feature, used in other studies as a binning characteristic of IRS spectra \citep[e.g.][]{Lambrides2019}, so our AGN fractions are independent of the atomic features we analyze in this paper. We aim to evenly span the range of AGN fractions from 0-1 with roughly equal number of galaxies per bin. Due to lower numbers of strong AGN sources, we end up with $\sim40$ galaxies in each of the lowest AGN fraction bins and 20~-~25 galaxies in the two highest AGN fraction bins (Table \ref{tab:bins}). The average properties of the galaxies in each bin including AGN fractions, L$_{\rm{IR}}$, luminosity distances, and merger stages are plotted in Figure \ref{fig:bincharacteristics} and the median values are listed in Table \ref{tab:bins}. Our six bins have similar distributions of L$_{\rm{IR}}$ and distance, except Bin 6 (the highest AGN fraction bin) where the sources are shifted to slightly higher L$_{\rm{IR}}$ and distance. However, the offset in median distance is small compared to the range of distance within each bin. While the galaxies in Bin 6 are slightly more distant and therefore slightly more luminous, we note that the distribution of L$_{\rm{IR,SF}}$, the portion of IR luminosity from star formation (dashed distributions in panel 2 of Figure \ref{fig:bincharacteristics}), is consistent across the six bins.

\begin{deluxetable*}{cccccc}
\tablenum{1}
\tablecaption{Properties of our six mid-IR AGN fraction bins \label{tab:bins}}
\tablewidth{0pt}
\tablehead{
\colhead{Bin} & \colhead{Number} & \colhead{Median (low, high)} & \colhead{} & \colhead{Median (Q1, Q3)} & \colhead{Median (Q1, Q3)} \\
\colhead{} & \colhead{of} & \colhead{mid-IR} & \colhead{Median (Q1, Q3)} & \colhead{Luminosity} & \colhead{SH-LH} \\
\colhead{} & \colhead{Galaxies} & \colhead{AGN Fraction} & \colhead{L$_{\mathrm{IR}}$} & \colhead{Distance (Mpc)} & \colhead{Scale Factor}
}
\decimalcolnumbers
\startdata
Bin 1 & 39 & 0.11 (0, 0.15) & 11.25 (11.1, 11.5) & 86.0 (76.3, 112.5) & 1.5 (1.3, 1.6)\\
Bin 2 & 38 & 0.19 (0.16, 0.23) & 11.25 (11.1, 11.5) & 89.7 (69.9, 129.5) & 1.6 (1.3, 1.8)\\
Bin 3 & 41 & 0.28 (0.24, 0.34) & 11.4 (11.2, 11.6) & 92.1 (72.3, 120.5) & 1.3 (1.2, 1.6)\\
Bin 4 & 41 & 0.40 (0.35, 0.56) & 11.4 (11.1, 11.7) & 106.0 (77.2, 155.0) & 1.6 (1.2, 1.9)\\
Bin 5 & 20 & 0.61 (0.56, 0.80) & 11.4 (11.1, 11.7) & 111.0 (86.4, 182.0) & 1.2 (1.2, 1.6)\\
Bin 6 & 24 & 0.96 (0.81, 1.0) & 11.8 (11.4, 12.0) & 148.5 (87.3, 197.3) & 1.2 (1.1, 1.3)\\
\enddata
\tablecomments{(3) Measured from the 6.2 $\mu$m PAH EQW \citep{DiazSantos2017}.}
\end{deluxetable*}

\subsection{Mid-IR spectral stacking}
\label{sec:stacking}
\begin{figure}[ht]
    \centering
    \includegraphics[width=7cm]{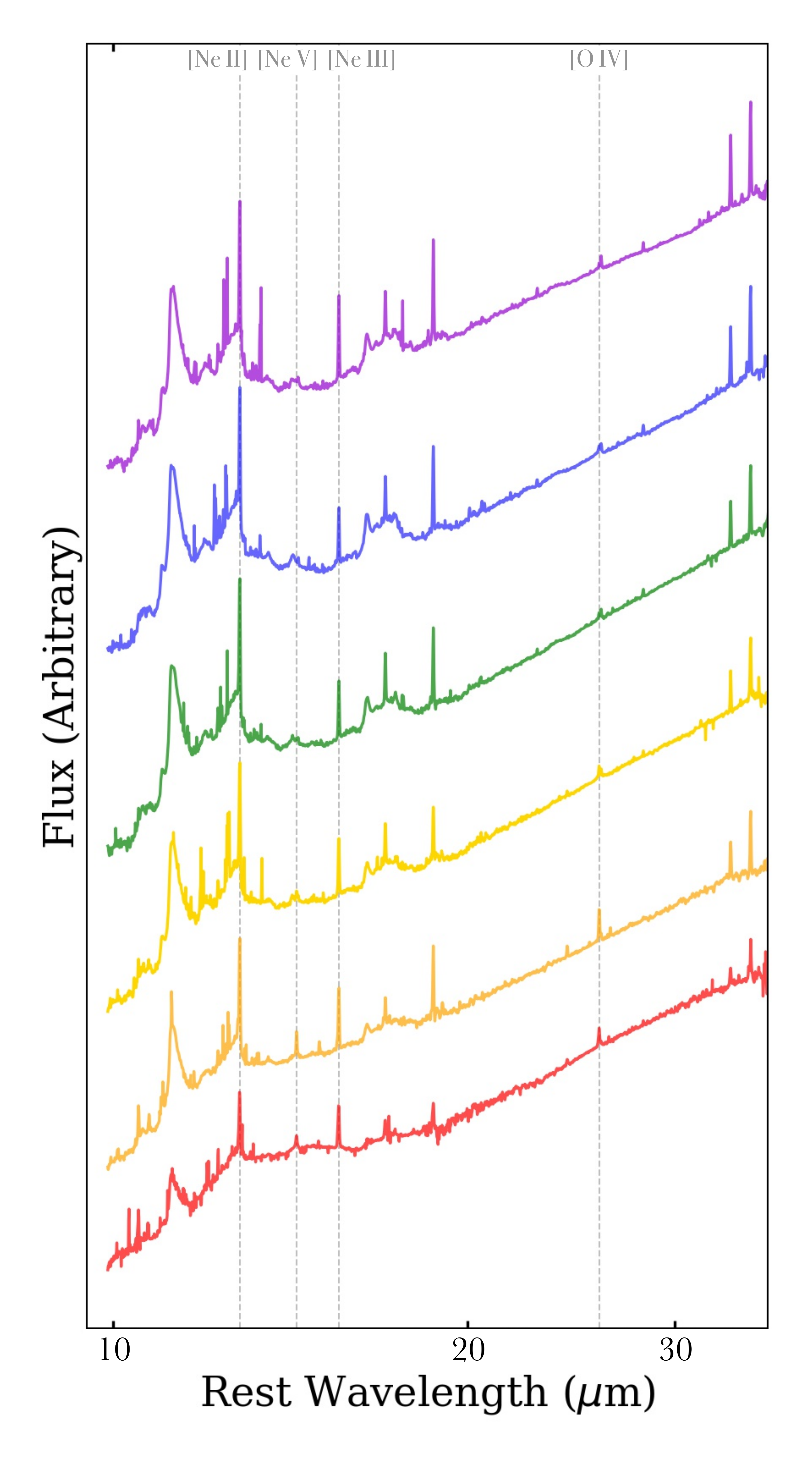}
    \caption{Stacked {\it Spitzer}/IRS spectra for our 6 bins of AGN fraction (increasing from top to bottom, same color code as Figure \ref{fig:bincharacteristics}) offset vertically for clarity. In this figure, the SH spectrum (9.9-19.6 $\mu$m observed) has been scaled up to match the LH spectrum (18.9-36.9 $\mu$m observed) before being shifted to the rest-frame. The main lines studied in this paper are denoted by the dashed gray lines with labels at the top. 
    }
    \label{fig:stackcomparison}
\end{figure}

For each galaxy, we combine the {\it Spitzer}/IRS SH and LH spectral channels and align them in the rest-frame using redshifts from \cite{Armus2009}. In our sample, [Ne~II], [Ne~V], and [Ne~III] fall in the SH slit (width $\sim 4.7"$), while [O~IV] lies in the LH slit, which is roughly twice as large (width $\sim 11.1"$). For the median redshifts in each bin of AGN fraction (Figure \ref{fig:bincharacteristics}), the SH (LH) slit spans a range in physical size of $1.8-3.1$ ($4.2-7.3$) kpc between the first and last bin. Owing to the mismatch in aperture size, the raw IRS spectra for each source shows a discontinuity in flux between the channels. Using the overlap regions between the two, we calculate the scale factor to bring the SH spectrum in line with the LH spectrum. The scale factors range from 1.2~--~1.6 (Table \ref{tab:bins}), and show only a very weak dependence on the AGN fraction (the highest AGN fraction bins have slightly lower scale factors, as expected since more of the emission is from the nuclear region and the bin is slightly more distant on average). In Figure \ref{fig:stackcomparison}, we scale the SH spectrum up for visual clarity using the measured scale factors. 

Approximately 40 sources spread evenly across the six bins are flagged by \cite{Inami2013} as being blended with additional sources in the LH slit. We examined the effects of excluding these blended sources from the stacks. When these 40 sources are not included, the noise in the stacked spectra increases as expected when stacking fewer sources, but the measured line fluxes are effectively unchanged from the stacked spectra where the blended sources are included. The stacked spectra and corresponding line measurements in this paper are based on the full sample of 203 GOALS sources.

Each spectrum in each bin is normalized to ensure the resulting stack is not dominated by intrinsically more luminous or nearby galaxies and to reduce the scatter due to these intrinsic differences in flux. We normalize at $\sim25$  $\mu$m, where the continuum is relatively smooth and free from major lines to provide the cleanest stacked spectra. Shortward of $\sim20$ $\mu$m, the continuum is littered with features and normalizing to the continuum there is difficult. We tested our choice of normalization by re-doing our stacking and normalizing at $\sim15\,\mu$m, between the [Ne~V] and [Ne~III] lines. Our results do not change qualitatively with the shorter-wavelength normalization but the line measurements are noisier. Specifically, the measurements of [O~IV] are noisier with a normalization further away from the line, and all line measurements in the AGN-dominated Bin 6 have lower SNRs due to the noisy continuum shortward of $\sim20$ $\mu$m. All spectra are interpolated onto a common wavelength array taken from NGC1068, the nearest galaxy in the sample with the coarsest rest-frame spectral resolution.

To create the stacked spectra for each bin, we calculate the mean of the fluxes at each wavelength pixel in the spectrum and calculate an error as the standard error of the fluxes. We also created the stacked spectra using the median and found they were consistent with the mean stacks within the uncertainties. The difference between the line measurements from the mean and median stacked spectra is on average 14\%, and [Ne~V] is not as well-detected in the median stacks. We restore the average intrinsic luminosity of the sources in each bin by re-normalizing the stacked spectrum to the median flux of the input galaxies in that bin. Our final six stacked spectra are shown in Figure \ref{fig:stackcomparison} and are included as a machine-readable table.

\subsection{Spectral line fitting}

The stacked spectra (Figure \ref{fig:stackcomparison}) show a number of features, including the broad PAH lines, as well as [Ne~II], [Ne~III], [Ne~V], and [O~IV]. We focus our analysis on these four fine-structure lines. The line fluxes are measured by anchoring a linear continuum on either side of the line and fitting a Gaussian.
When multiple lines were blended together ([Ne~V] with [Cl~II] and [O~IV] with [Fe~II]), we fit both lines simultaneously. For [Ne~II], [Ne~III], and [O~IV], we fit the amplitude, central wavelength, and width; for [Ne~V], which is located on top of a PAH dust feature and very faint in certain bins, we fix the width to 0.013 $\mu$m and fit only the central wavelength\footnote{We explored fixing the central wavelength but found this was too restrictive given the blending of [Ne~V] with [Cl~II].} and amplitude. 

\begin{figure*}[ht]
    \centering
    \includegraphics[width=16.5cm]{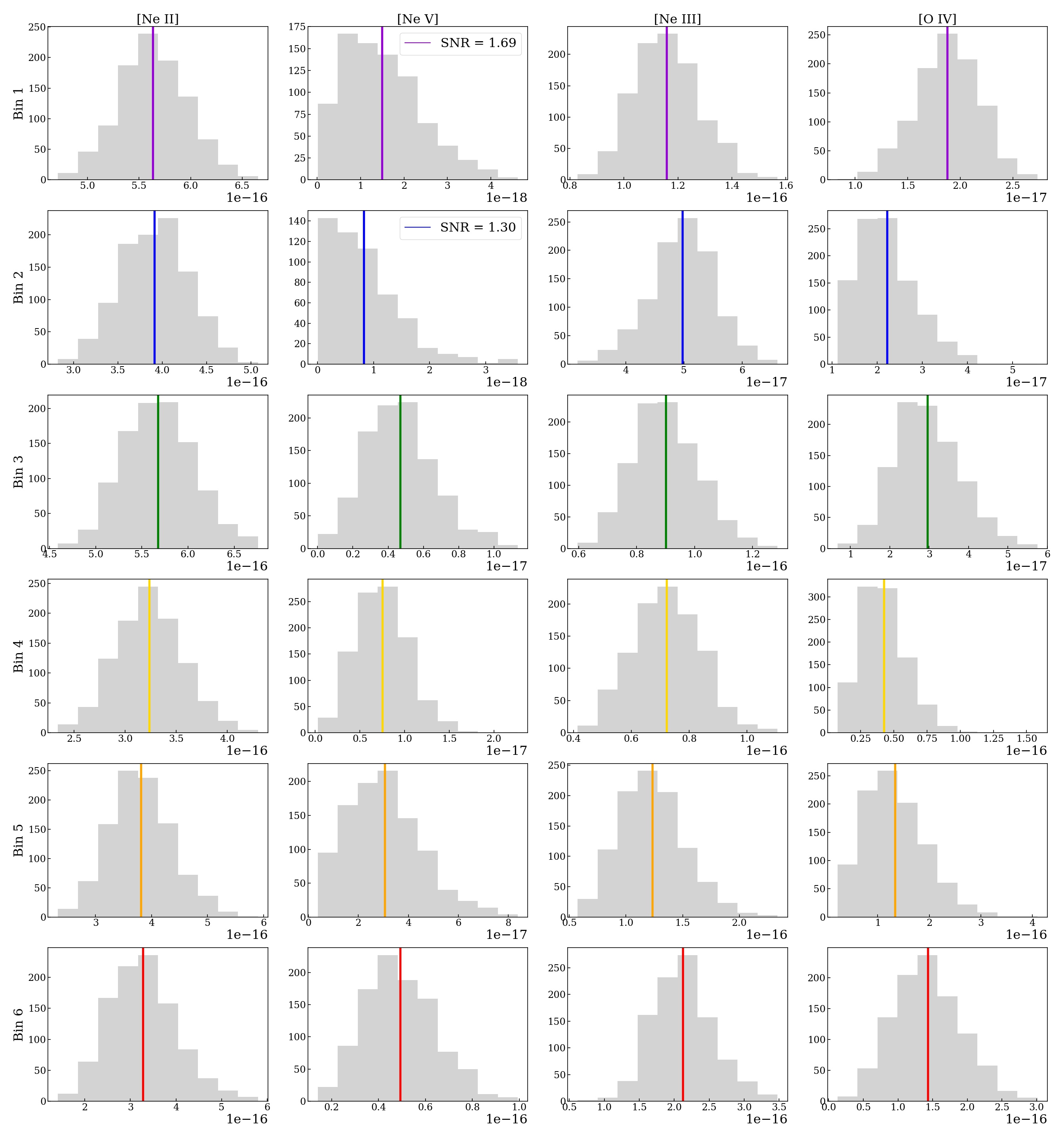}
    \caption{Distribution of the measured bootstrapped fluxes for each bin (rows) and each line (columns), with the median of the distribution as a colored line. The lines are detected in every bin except for [Ne~V] in the first two bins, where the SNR is given on the plot. In these two bins, we report a $3\sigma$ upper limit on the luminosity of [Ne~V] in all subsequent analysis. 
    }
    \label{fig:luminositydistribution}
\end{figure*}

We derive errors on the line fluxes by performing a bootstrap analysis to account for variation within the bins. For each bin, consisting of N galaxies, we draw N random galaxies with replacement, stack their spectra according to the procedure in Section \ref{sec:stacking}, and fit the lines. Repeating this procedure 1000 times for each bin produces a distribution of fluxes for each line in each bin (Figure \ref{fig:luminositydistribution}). These distributions are well-behaved; we take the measured flux of the line from the stack as our final line flux, and use the standard deviation of the bootstrap distribution for the error. The mean and median are very consistent for each distribution, and the mean of the distribution is very similar to the measured line flux from the original stacked spectrum except for [Ne~V] in Bins 1 and 2, where the line is not detected in most bootstrap trials.
The [Ne~II], [Ne~III], and [O~IV] lines are detected at greater than 2$\sigma$ significance in all bins (line flux is $>~2\times$ the bootstrap uncertainty). [Ne~V], the faintest line studied, does not show a Gaussian distribution in the two lowest AGN fraction bins and the mean line fluxes are not significant given the width of the distributions. We therefore consider [Ne~V] to be undetected in these two bins and calculate the $3\sigma$ upper limit on the flux as three times the bootstrapped uncertainty.
Finally, we convert our fluxes to luminosities using the median luminosity distance of the galaxies in each bin (Table \ref{tab:bins}). Our measured line fluxes and uncertainties are given in Table \ref{tab:fluxes}.

\begin{deluxetable}{ccccc}
\tablenum{2}
\tablecaption{Measured line fluxes in the stacked spectra for each AGN fraction bin. These are the raw fluxes as measured through the SH and LH slits and the uncertainties come from the bootstrapping the sample. The [Ne V] line in bins 1 and 2 is undetected ($<2\sigma$) and so we report $3\sigma$ upper limits. \label{tab:fluxes}}
\tablewidth{0pt}
\tablehead{
\colhead{Bin} & \colhead{[Ne~II]} & \colhead{[Ne~V]} & \colhead{[Ne~III]} & \colhead{[O~IV]}\\
\colhead{} & \colhead{$10^{-16}$ W/m$^2$} & \colhead{$10^{-18}$ W/m$^2$} & \colhead{$10^{-17}$ W/m$^2$} & \colhead{$10^{-17}$ W/m$^2$}
}
\decimalcolnumbers
\startdata
Bin 1 & 5.61 $\pm$ 0.33 & $<$ 10.4 & 11.6 $\pm$ 1.2 & 1.86 $\pm$ 0.29 \\
Bin 2 & 3.89 $\pm$ 0.38 & $<$ 7.29 & 4.94 $\pm$ 0.55 & 2.23 $\pm$ 0.65 \\
Bin 3 & 5.68 $\pm$ 0.39 & 4.77 $\pm$ 1.98 & 9.04 $\pm$ 1.13 & 2.99 $\pm$ 0.84 \\
Bin 4 & 3.24 $\pm$ 0.32 & 7.48 $\pm$ 2.82 & 7.23 $\pm$ 1.11 & 4.28 $\pm$ 1.67 \\
Bin 5 & 3.81 $\pm$ 0.54 & 29.7 $\pm$ 14.9 & 12.1 $\pm$ 2.7 & 13.0 $\pm$ 6.1 \\
Bin 6 & 3.26 $\pm$ 0.75 & 49.4 $\pm$ 14.9 & 20.9 $\pm$ 4.3 & 14.5 $\pm$ 4.7 \\
\enddata
\end{deluxetable}

\subsection{Cross validation}
\label{sec:consistency}

Given our stacked spectra (Figure \ref{fig:stackcomparison}) and line measurements (Figure \ref{fig:luminositydistribution}), we now verify our measurements with several tests. 

\begin{figure*}
    \centering
    \includegraphics[width=14cm]{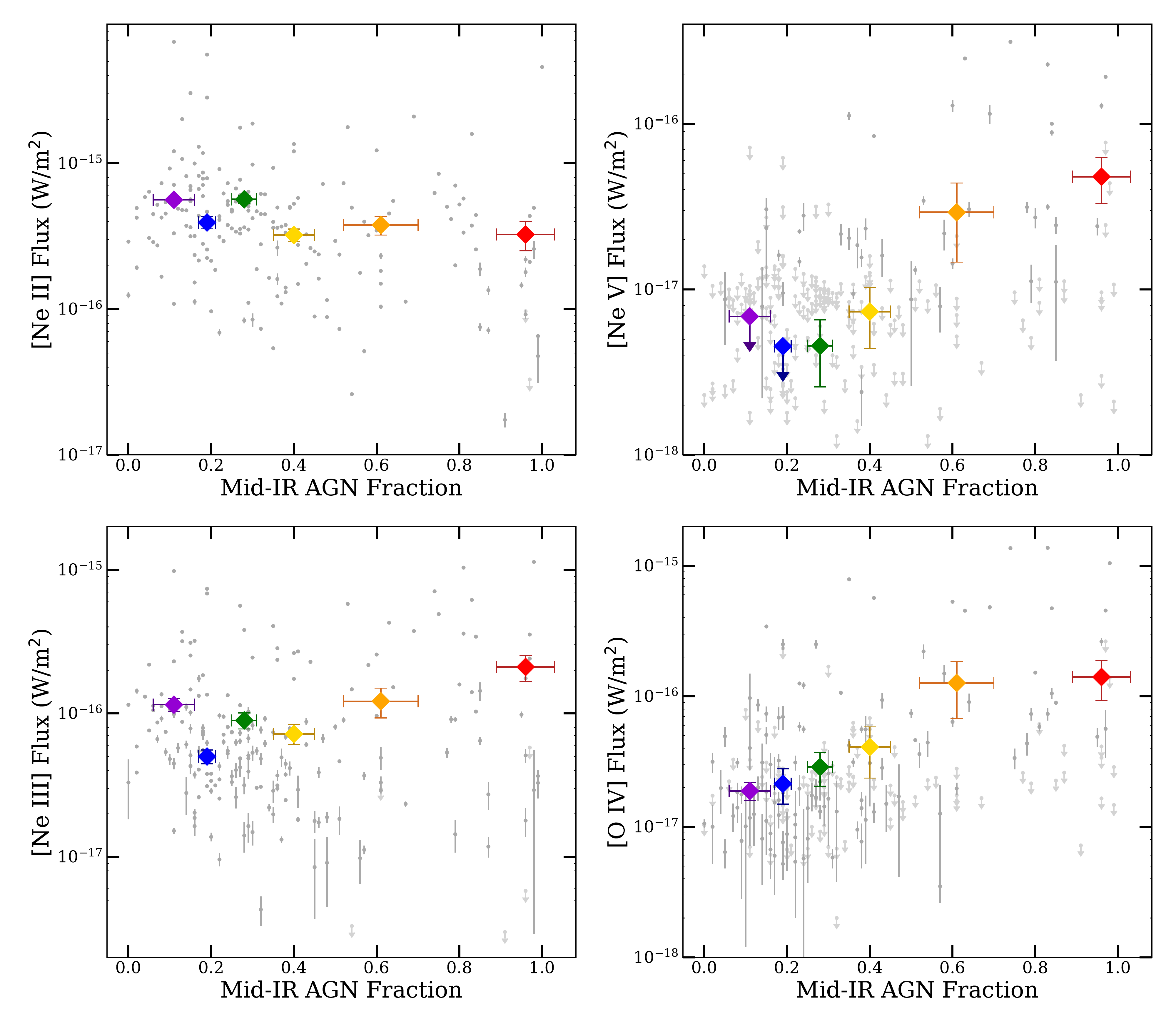}
    \caption{Line fluxes for [Ne~II], [Ne~III], [Ne~V] and [O~IV] as a function of AGN fraction from our stacked spectra (colored) compared to individual galaxy measurements (darker gray) and upper limits (lighter gray) from \cite{Inami2013}. The x-errors are the standard deviation of the AGN fractions in each bin. While [Ne~V] is only detected in 18\% of the individual sources and [O~IV] is detected in just over half of individual sources, the stacking analysis permits detection of [O~IV] in all bins and [Ne~V] down to AGN fraction of 0.25. 
    }
    \label{fig:LinesVsInami}
\end{figure*}

As a consistency check on our measured line luminosities, we compare the fluxes measured from our stacked spectra to the values for individual GOALS galaxies measured in \cite{Inami2013}. The four panels of Figure \ref{fig:LinesVsInami} show our measured stacked fluxes (colored symbols) on top of the measurements and upper limits for the individual galaxies from \cite{Inami2013} (gray symbols). [Ne~II] and [Ne~III] are detected in almost all individual sources (98\% and 96\%, respectively) and our stacked measurements represent the average of these detections. Since [O~IV] and [Ne~V] are only detected in a subset of GOALS galaxies (55\% and 18\%, respectively), our stacked measurements can provide a better estimate of the true average for these lines in each AGN fraction bin. 

Next, we explore the implications of the limited size of the IRS slit relative to the size of the GOALS galaxies. Since the IRS slit might not capture all the emission from each galaxy, it is important to ensure that this does not introduce any biases that may affect our results. We checked this in two ways. First, we observe that there is no significant difference in the distribution of distances between the bins (see Figure \ref{fig:bincharacteristics}, third panel from left, and the quartile ranges in Table \ref{tab:bins}, column 4). Bin 6 has a slightly higher median distance, but the difference is not significant given the width of the distribution. We also examine the distribution of intrinsic mid-IR core sizes of the GOALS galaxies, calculated from the observed mid-IR core sizes from IRS reported in \cite{DiazSantos2011}. The distribution of core sizes is very consistent across the bins, with a median size of 2.5 kpc (with quartiles at 1.7 and 3.6 kpc). Therefore, on average, the physical scales traced by the superimposed slits are independent of AGN fraction.

\begin{figure}[ht]
    \centering
    \includegraphics[width=7cm]{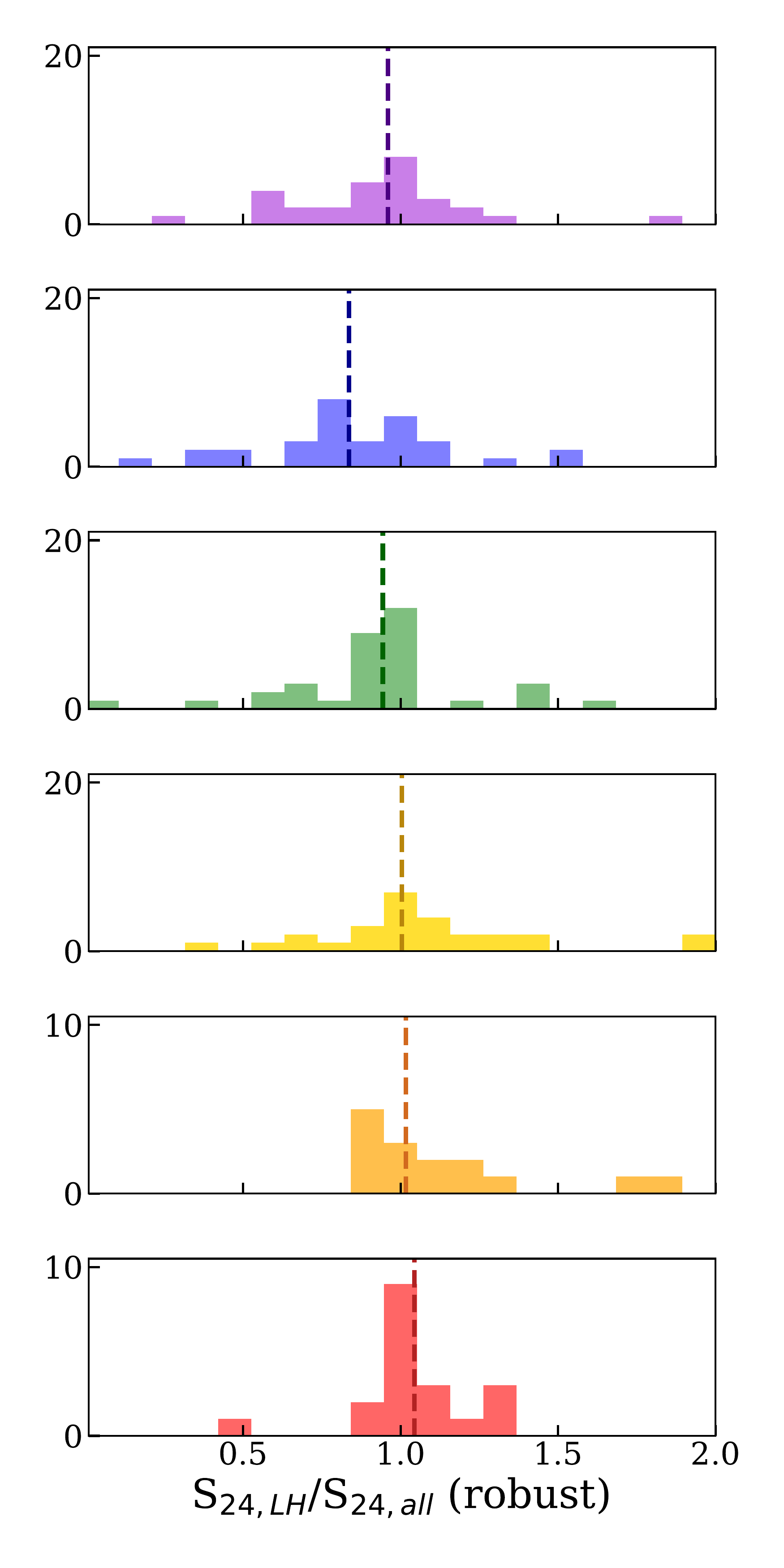}
    \caption{Distribution of $S_{24,\rm{LH}}/S_{24,\rm{all}}$ for each AGN fraction bin. $S_{24,\rm{LH}}$ is the synthetic flux measured on the LH IRS spectrum and $S_{24,\rm{all}}$ is the total MIPS flux. The vertical dashed line is the median of the distribution. While there appears to be a slight increase in the median ratio with AGN fraction, the distributions are broad and this trend is not significant.}
    \label{fig:s24comparison}
\end{figure}

Another concern is that, because AGN emission is centrally concentrated within galaxies, while emission from star formation is distributed throughout, the slit will capture almost all of the [Ne V] emission while only capturing part of the [Ne II] emission of the whole galaxy, thus introducing an artificial bias in the relative ratio. To check for this effect, we examined the ratio of the galaxy-integrated MIPS 24 $\mu$m flux densities ($S_{24,\rm{all}}$), and \textit{synthetic} MIPS 24 $\mu$m photometry created by applying the MIPS 24 $\mu$m filter to the Long-High (LH) IRS spectrum ($S_{24,\rm{LH}}$). The distribution of the ratio $S_{24,\rm{LH}}/S_{24,\rm{all}}$ for each bin is plotted in Figure \ref{fig:s24comparison}. We have excluded any sources where there is blending in the LH slit.
The median ratio $S_{24,\rm{LH}}/S_{24,\rm{all}}$ is nearly constant over all AGN fraction bins, only slightly lower in Bin 2. Similarly, \cite{Stierwalt2014} measured the ratio of the 8 $\mu$m flux integrated over the whole galaxy relative to the flux captured by the Short-Low slit, which we confirm to be constant across our AGN fraction bins as well. For sources dominated by the AGN, we can assume the 24 $\mu$m emission is hot dust from the AGN and will trace a similar extent as the [Ne V] emission, and that the 8 $\mu$m emission traces the star-formation and is thus comparable in extent to [Ne II]. We conclude that there are no obvious systematic biases in the [Ne V]/[Ne II] ratio as a function of AGN fraction in terms of how much of either emission line is captured by the IRS slit across the sample.

When measuring AGN line luminosities like [Ne~V], is not correct to use the SH spectrum that has been scaled up to meet the LH spectrum, because the [Ne~V] emission will be centrally concentrated and should already be captured in the SH slit. However, the larger LH slit can cover more star formation in these galaxies and so it may be reasonable to use the scaled SH spectrum to measure the [Ne~II] emission. We therefore consider the effects of scaling the [Ne~II] flux up to match the LH spectrum using the median scale factors listed in Table \ref{tab:bins} when examining the line ratios and fits discussed in Sections \ref{sec:results} and \ref{sec:discussion}.

\begin{figure}[ht]
    \centering
    \includegraphics[width=8.33cm]{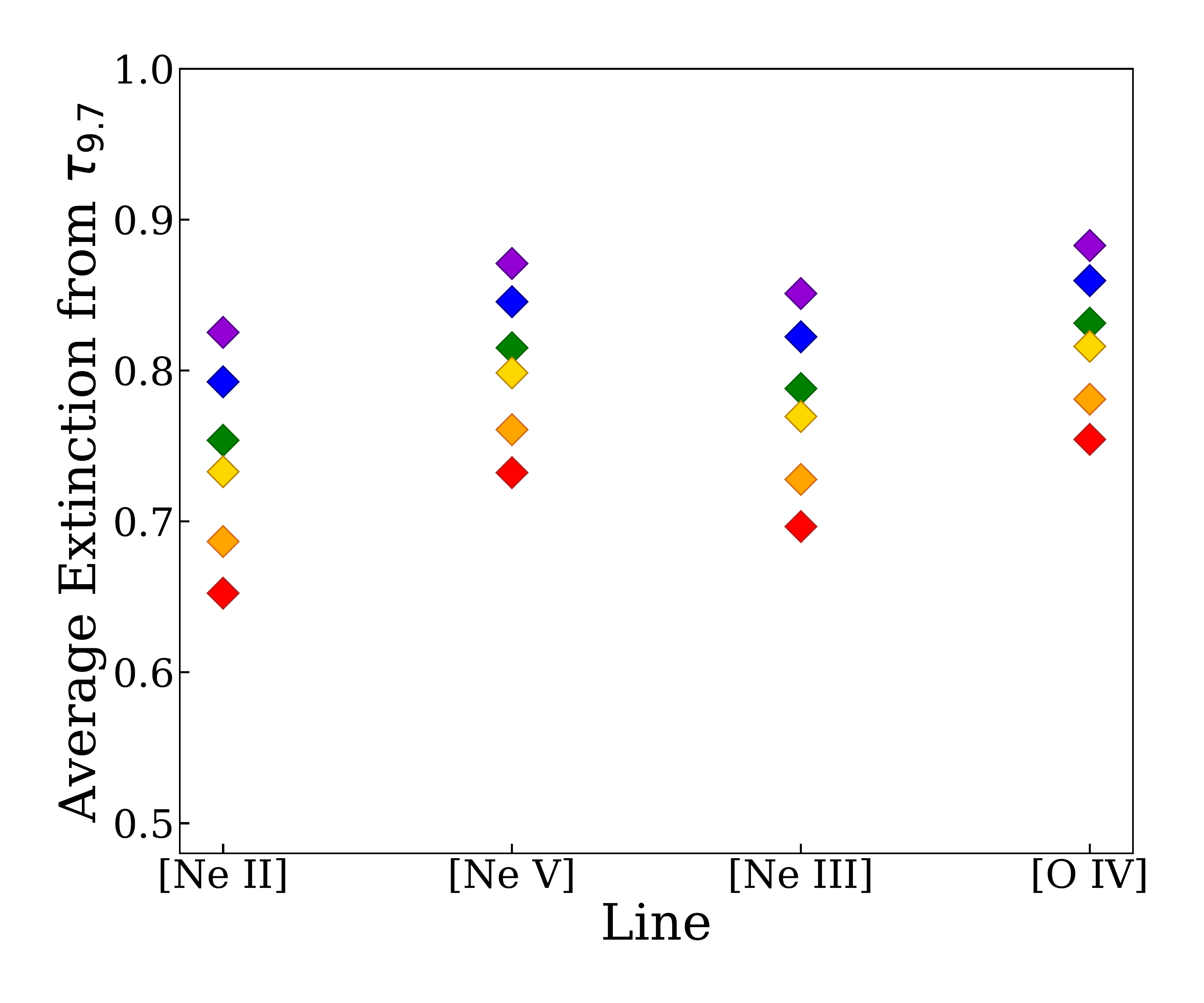}
    \caption{Dust extinction (fraction of emitted light observed) implied for each line and for each AGN fraction bin estimated from the $\tau_{9.7}$ and the \cite{Draine2001} dust extinction model.
    }
    \label{fig:extinction}
\end{figure}

Finally, we must ensure that dust extinction does not preferentially affect some lines and/or some AGN fraction bins more than others. The optical depth at 9.7$\,\mu$m, $\tau_{9.7}$, has been measured for each galaxy in the GOALS sample \citep{Stierwalt2013}. We find the distribution of $\tau_{9.7}$ is consistent across our AGN fraction bins. Similarly, \cite{Stierwalt2013} found that while there were not many high $\tau_{9.7}$ sources with high PAH EQW (which implies low AGN fraction), there is no strong relation between $\tau_{9.7}$ and PAH EQW. 
We further examine the dust extinction on the lines by utilizing the \cite{Draine2001} dust extinction models. These models allow us to use the $\tau_{9.7}$ of each galaxy to calculate the implied average extinction of each bin at the wavelength of each line. Figure \ref{fig:extinction} shows the estimated extinction for each line and for each AGN fraction bin. This figure shows that the effect of extinction on any of the line ratios will be negligible, as the range of extinctions between lines in a single bin is very small ($\lesssim10$\%).
For line luminosities, we do see higher levels of dust extinction in the highest AGN fraction bins, with the differences in extinction corrections across the bins up to $\sim25$\%. However, our measured luminosities of [Ne~V] and [O~IV] increase by more than an order of magnitude from Bin 1 to Bin 6 (see Figure \ref{fig:alllines}), so any correction for dust extinction will not affect our results. 
\cite{Farrah2007} calculate the extinction on the mid-IR lines using the measured $A_{V}$ for a subset of local ULIRGs and find similarly that the extinction on the line ratios is negligible. Given these tests and to be consistent with other studies of these lines in the literature, we elect not to correct the lines for dust extinction and note that a correction for extinction would not significantly affect our results.

\section{Results} \label{sec:results}

\subsection{Line luminosities as a function of AGN fraction}

\begin{figure}[ht]
    \centering
    \includegraphics[width=8.33cm]{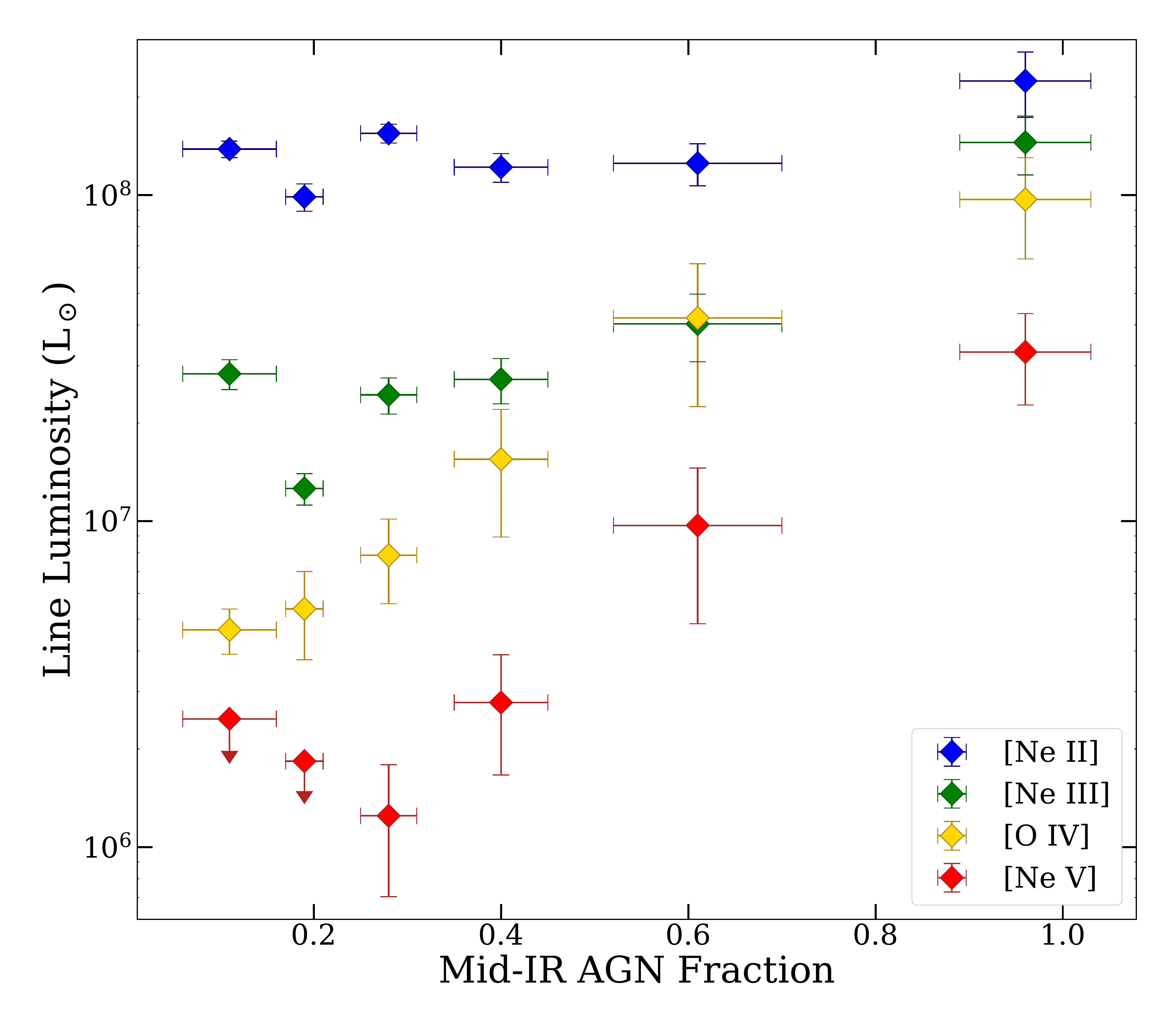}
    \caption{Measured line luminosities as a function of AGN fraction. Here we plot the intrinsic line luminosities of the galaxies contained in each bin as measured directly from the SH and LH stacked spectra (with no aperture scaling applied). The error on the AGN fraction is the standard deviation of the values in the bin, and is not shown on future plots. 
    }
    \label{fig:alllines}
\end{figure}

The luminosities of [Ne~II], [Ne~III], [Ne~V], and [O~IV] are plotted in Figure \ref{fig:alllines} as a function of AGN fraction. The errors on the AGN fraction are the standard deviations of the mid-IR AGN fractions in each bin: for visual clarity, these are not included on most subsequent plots. The luminosities of [Ne~V] and [O~IV] increase strongly with AGN fraction as expected. The [Ne~II] luminosity, on the other hand, is fairly consistent across the range of AGN fraction. The [Ne~III] luminosity is also fairly flat with respect to the AGN fraction, with a slight increase in the highest AGN fraction bin. We note that the luminosity of [Ne~III] decreases from Bin 1 to Bin 2; this is visible in the stacks as well as the bootstrap distribution. [Ne~II] also shows a similar effect to a lesser degree, which suggests it could be related to the intrinsic properties of the galaxies in the bin, although there is no indication of this in the distributions of $L_{\mathrm{IR}}$ in Figure \ref{fig:bincharacteristics}. Bin 2 does have a noticeably low median $S_{24}$ ratio (see Figure \ref{fig:s24comparison}), but correcting the line fluxes for the variations in $S_{24}$ between bins does not bring the Bin 2 [Ne~III] flux up to the luminosities found in Bins 1 and 3. Regardless, this change in luminosity of [Ne~III] between Bins 1 and 2 is much less significant than the strong and continuous increase in [O~IV] and [Ne~V] as a function of AGN fraction. The values of $\log(L\mathrm{[O\;IV]}/L_{\mathrm{IR}})$ for the six bins (in order of increasing AGN fraction) are -4.63, -4.56, -4.47, -4.24, -3.87, and -3.75.

\subsection{Line Ratios}
\begin{figure*}[ht]
    \centering
    \includegraphics[width=16.5cm]{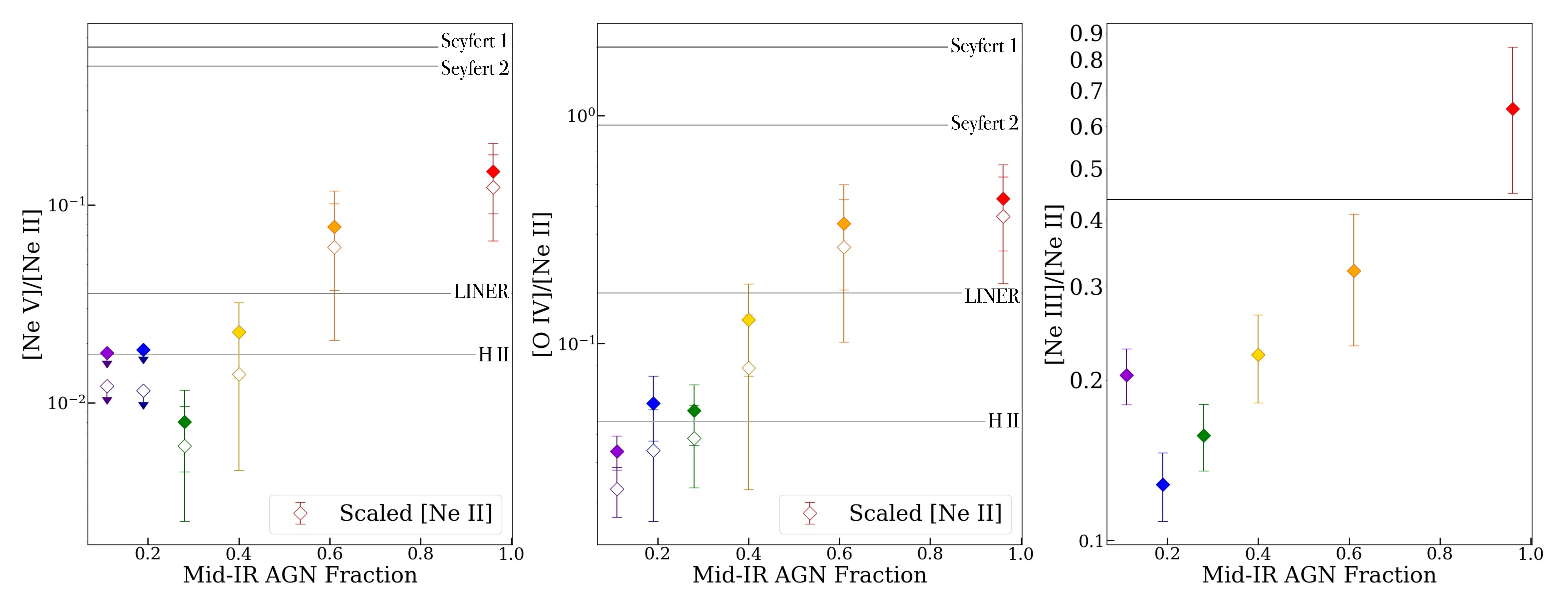}
    \caption{({\it left}) [Ne~V]/[Ne~II] and ({\it middle}) [O~IV]/[Ne~II] as a function of AGN fraction. 
    The filled symbols are calculated with the [Ne~II] flux as measured through the SH slit, which are consistent with the open symbols that include the [Ne~II] flux scaled to meet the LH slit. 
    Horizontal lines show the median ratios for different types of galaxies from \cite{PereiraSantaella2010}. ({\it right}) [Ne~III]/[Ne~II] with horizontal line showing the median line ratio for star forming galaxies from \cite{Ho2007}. The \cite{Ho2007} sample covers a much wider range of masses and extends to lower metallicities and so it is not surprising that the [Ne~III]/[Ne~II] ratios for our sample of LIRGs tend to be lower. 
    }
    \label{fig:lineratios}
\end{figure*}

As [Ne~II] is expected to trace the SFR \citep{Ho2007}, taking the ratio of the lines relative to [Ne~II] attempts to normalize out any dependence of the bins on SFR \citep[e.g.][]{Genzel1998,Sturm2002,Armus2004,Armus2007}. 
In Figure \ref{fig:lineratios}, we examine the line ratios [Ne~V]/[Ne~II] (left), [O~IV]/[Ne~II] (center), and [Ne~III]/[Ne~II] (right) as a function of AGN fraction. We show the ratios for [Ne~II] measured directly from the SH slit and when the [Ne~II] luminosity is scaled to match the LH slit using the scale factors listed in Table \ref{tab:bins}. We find that the ratios are consistent within their error bars whether or not [Ne~II] is scaled, as the scale factors across bins only range from 1.2--1.6.

All three ratios in Figure \ref{fig:lineratios} increase as the relative AGN power increases, with both [Ne~V]/[Ne~II] and [O~IV]/[Ne~II] increasing by over an order of magnitude. [O~IV]/[Ne~II] increases steadily from AGN fractions of 0.3--1 but appears somewhat constant below an AGN fraction of $\sim0.3$. In the left and center panels, we overplot the median ratios for different galaxy classes from the sample of LIRGs, QSOs, Seyfert galaxies, LINERs, and star-forming galaxies examined by \cite{PereiraSantaella2010} as horizontal lines. We find our lower-AGN fraction bins ($\lesssim0.3$) are consistent with line ratios found for H~II galaxies, while the higher AGN fraction bins are more consistent with the line ratios found between LINER and Seyfert 2 systems. This is roughly as expected---our LIRG bins all have substantial contributions from star formation even at the highest AGN fractions, and will not resemble pure AGN. In the right panel, we overplot the median [Ne~III]/[Ne~II] ratio for the sample of star-forming galaxies in \cite{Ho2007} as a horizontal line. The GOALS galaxies have less [Ne~III]/[Ne~II] emission than the \cite{Ho2007} sample except in the very highest AGN fraction bin. This is likely because the GOALS sample is limited to high-mass, and therefore high-metallicity, galaxies while the \cite{Ho2007} sample covers several orders of magnitude in $L_{\mathrm{IR}}$ and extends down to include low-metallicity dwarf galaxies.

\begin{figure}
    \centering
    \includegraphics[width=8.33cm]{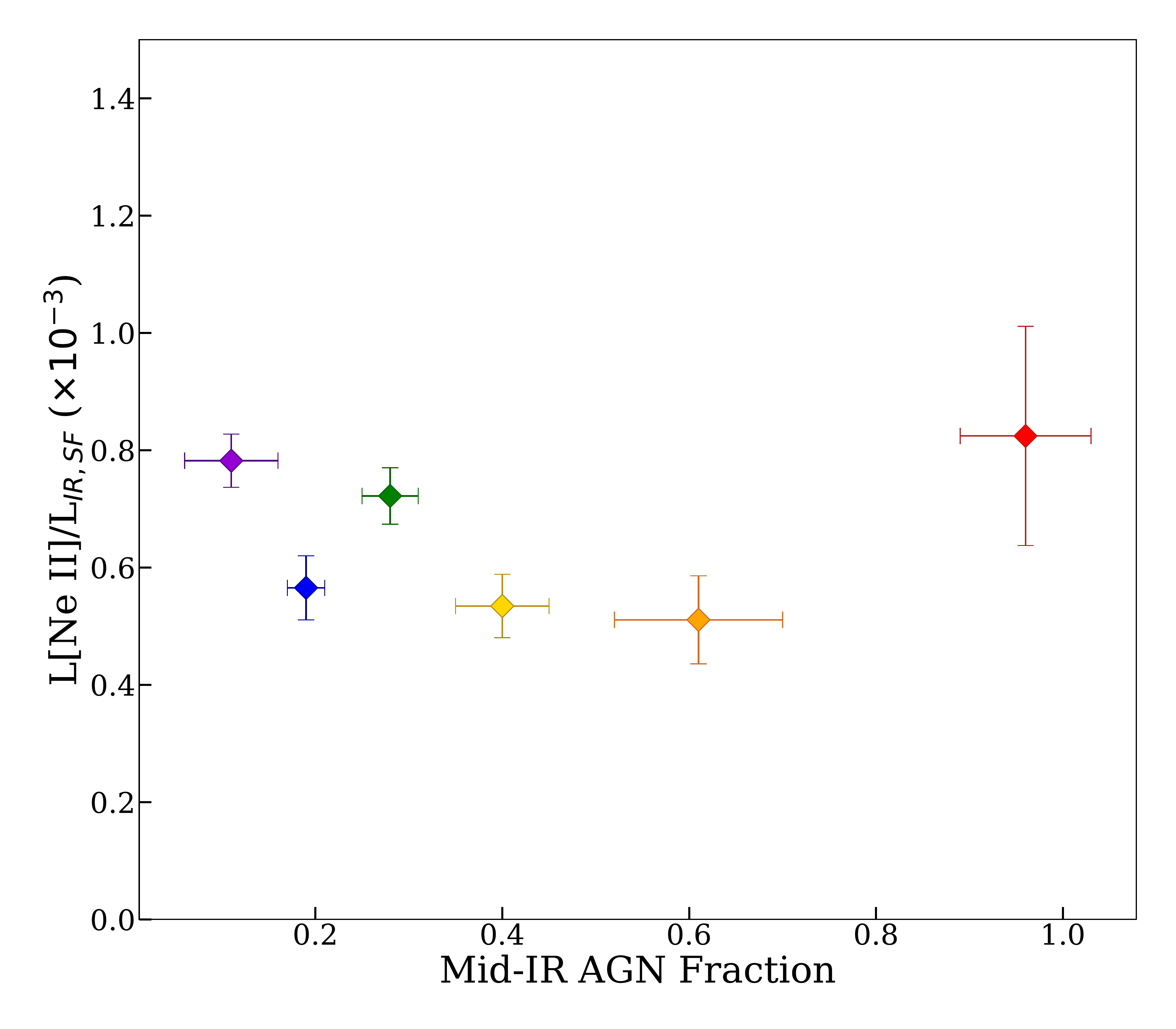}
    \caption{The ratio [Ne II]/L$_{IR,SF}$ as a function of AGN fraction for our six bins. We do not observe any evolution of this ratio with AGN fraction, though a larger [Ne II]/L$_{IR,SF}$ fraction could be hidden by the large uncertainty in Bin 6. We conclude the [Ne II] emission is dominated by star formation and can be used to determine [O IV]$_{SF}$ in Equation \ref{eq:oivwnev}.}
    \label{fig:neiilirsf}
\end{figure}

\subsection{Contribution of Star Formation and AGN to [O~IV]}
\label{sec:decompose}

\begin{figure*}[ht]
    \centering
    \includegraphics[width=16.5cm]{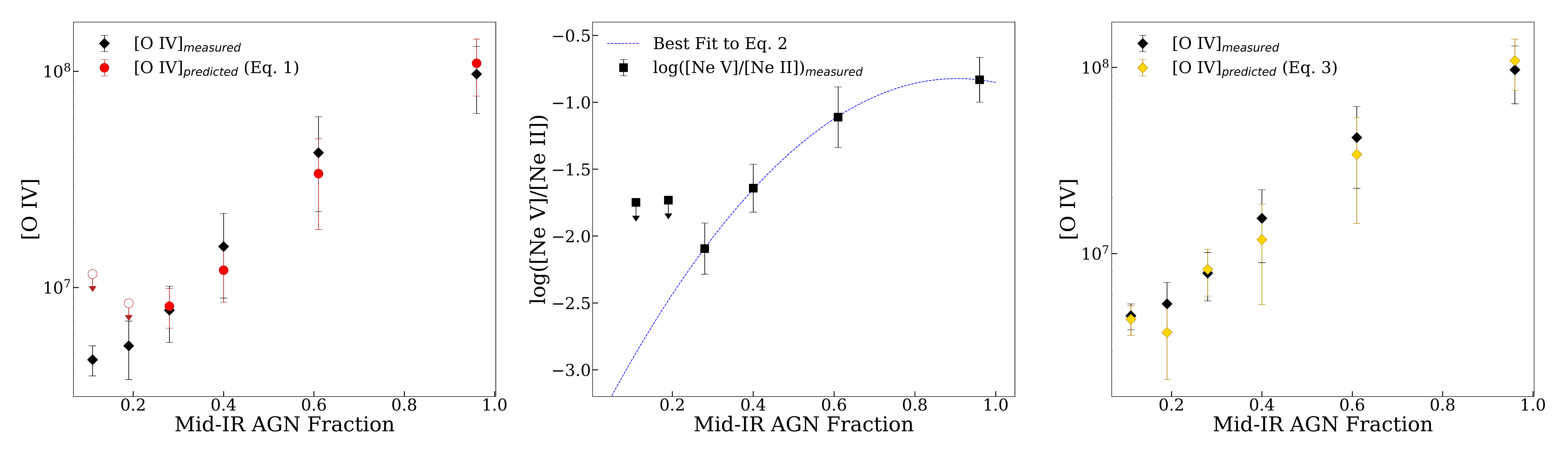}
    \caption{The process to predict the [O~IV] luminosity from [Ne~II] luminosity and AGN fraction. We fit [O~IV] as a function of [Ne~II] and [Ne~V] (left panel and Equation \ref{eq:oivwnev}) and fit log([Ne~V]/[Ne~II]) as a function of AGN fraction (center panel and Equation \ref{eq:ratio}. We then combine these two relations to predict [O~IV] as a function of [Ne~II] luminosity and AGN fraction (right panel). Our relation in terms of [Ne~II] and AGN fraction reproduces the [O~IV] measurements, including in Bins 1 and 2 where [Ne~V] is not detected. All [Ne~II] fluxes in this figure are unscaled.
    }
    \label{fig:fittingoiv}
\end{figure*}

In Figure \ref{fig:lineratios}, we observe that the [O~IV]/[Ne~II] line ratio appears to flatten out below an AGN fraction of 0.3. Keeping in mind the ionization potential of [O~IV], this could be because both [O~IV] and [Ne~II] are dominated by star formation in this regime. Figure \ref{fig:neiilirsf} confirms that the ratio [Ne~II]/L$_{\mathrm{IR,SF}}$ is relatively constant (between $5.1 \times 10^{-4}$ and $8.1 \times 10^{-4}$) and does not appear to evolve with AGN fraction, although the large uncertainty on Bin 6 could disguise $\sim30\%$ more [Ne~II] emission. This demonstrates that the [Ne~II] luminosity is most likely dominated by the starburst with minimal contribution from the AGN. Given that [Ne~V] is considered a pure AGN tracer even at low AGN fractions, and the [Ne~II] emission in our sample is dominated by the starburst, we decompose our measured [O~IV] lines into contributions from star formation and AGN. We fit a linear combination of [Ne~II] and [Ne~V] to the total [O~IV] luminosity:

\begin{equation} \label{eq:oivwnev}
    \mathrm{[O\:IV]} = \mathrm{[O\:IV]_{SF}}+\mathrm{[O\:IV]_{AGN}} = a\,\mathrm{[Ne\:II]}\:+\:b\,\mathrm{[Ne\:V]},
\end{equation}

and find best-fit coefficients of $a = (2.795\,\pm\,0.009)\times 10^{-2}$ and $b = 3.107\,\pm\,0.274$, with a average offset of 16\% between the data and fit for Bins 3-6, well within the uncertainties on the luminosities. Figure \ref{fig:fittingoiv} (left panel) shows the predicted total [O~IV] from Equation \ref{eq:oivwnev} (red) compared to our measured total [O~IV]. 

Our ability to determine the relative contributions of star formation and AGN accretion to the luminosity of [O~IV] is somewhat limited by the non-detections of [Ne~V] in our lowest two AGN fraction bins. However, as we found in Figure \ref{fig:lineratios}, [Ne~V]/[Ne~II] has a strong dependence on AGN fraction. This ratio is well-fit by a quadratic function\footnote{A linear fit to the [Ne~V]/[Ne~II] ratio yielded a significantly worse reduced $\chi^2$.} defined as: 

\begin{equation} \label{eq:ratio}
    \mathrm{log([Ne\:V]/[Ne\:II])} = c\,\mathrm{f_{AGN}}^2\:+\:d\,\mathrm{f_{AGN}}\:+\:e, 
\end{equation}

with best fit coefficients of $c = -3.246\,\pm\,0.014$, $d = 5.874\,\pm\,0.023$, and $e = -3.480\,\pm\,0.002$, with a average offset of 0.3\% between the data and the fit in Bins 2-6.
Figure \ref{fig:fittingoiv} (middle panel) shows the best-fit function compared to our measured [Ne~V]/[Ne~II]. 
We rearrange Equation \ref{eq:ratio} to isolate [Ne V] and substitute it into Equation \ref{eq:oivwnev}, yielding:

\begin{equation}\label{eq:oivpred}
    \mathrm{[O\:IV]_{pred}} = \mathrm{[Ne\:II]}(a\:+\:b\!\times\!10^{\,c\,\mathrm{f_{AGN}}^2\:+\:d\,\mathrm{f_{AGN}}\:+\:e})
\end{equation}

where $a$, $b$, $c$, $d$, and $e$ are the same as above. Although both terms now depend on [Ne~II], the first remains the contribution of star formation to [O~IV], and the second the AGN contribution. The uncertainties on all coefficients reflect the standard error on the median AGN fraction and uncertainties on the line fluxes. We explored the covariance of these coefficients and found that the errors on the coefficients are much less than the measurement uncertainty. We therefore propagate only the errors on the measured quantities and not on the fit coefficients into the uncertainties in the remainder of this paper.

These fits were performed on unscaled [Ne~II] fluxes. If the fit is performed using [Ne~II] fluxes scaled upward by the median scale factors in Table \ref{tab:bins}, the coefficients change; however, any properties derived from [O~IV]$_{AGN}$ (such as the black hole accretion rate) will remain the same, as the fit is calibrated to the total [O~IV] luminosity in both cases. We provide values of $a$, $b$, $c$, $d$, and $e$ for both unscaled and scaled [Ne~II] fluxes in Table \ref{tab:coefficients}.

\begin{deluxetable*}{cccccc}
\tablenum{3}
\tablecaption{Coefficients $a$, $b$, $c$, $d$, and $e$ from Equation \ref{eq:oivpred} for unscaled and scaled [Ne~II] fluxes, where the scale factors were calculated to align the SH and LH spectra (Table \ref{tab:bins}). Throughout the paper, we use the unscaled line fluxes. \label{tab:coefficients}}
\tablewidth{0pt}
\tablehead{
\colhead{} & \colhead{$a$} & \colhead{$b$} & \colhead{$c$} & \colhead{$d$} & \colhead{$e$}
}
\decimalcolnumbers
\startdata
Unscaled & $(2.795 \pm 0.009) \times 10^{-2}$ & $3.107 \pm 0.274 $ & $-3.246 \pm 0.014$ & $5.874 \pm 0.023$ & $-3.480 \pm 0.002$ \\
Scaled & $(2.127 \pm 0.004) \times 10^{-2}$ & $3.083 \pm 0.222$ & $-2.937 \pm 0.052$ & $5.606 \pm 0.086$ & $-3.582 \pm 0.006$ \\
\enddata
\end{deluxetable*}

This predicted [O~IV] luminosity from Equation \ref{eq:oivpred} is overplotted with the measured [O~IV] luminosity in the right panel of Figure \ref{fig:fittingoiv}, where the uncertainties on [O~IV]$\mathrm{_{predicted}}$ reflect the errors on the measured [O~IV] and [Ne~II] fluxes. Not only does [O~IV]$\mathrm{_{predicted}}$ match [O~IV]$\mathrm{_{measured}}$ in the upper four bins as expected, it fits the [O~IV]$\mathrm{_{measured}}$ in the lower two bins, where previously the [Ne~V] non-detections limited our ability to fit [O~IV]. As these fits are based on upper limits below an AGN fraction of $\sim 0.2$, they should be used with more caution at this lowest-AGN regime.

\begin{figure}
    \centering
    \includegraphics[width=8.33cm]{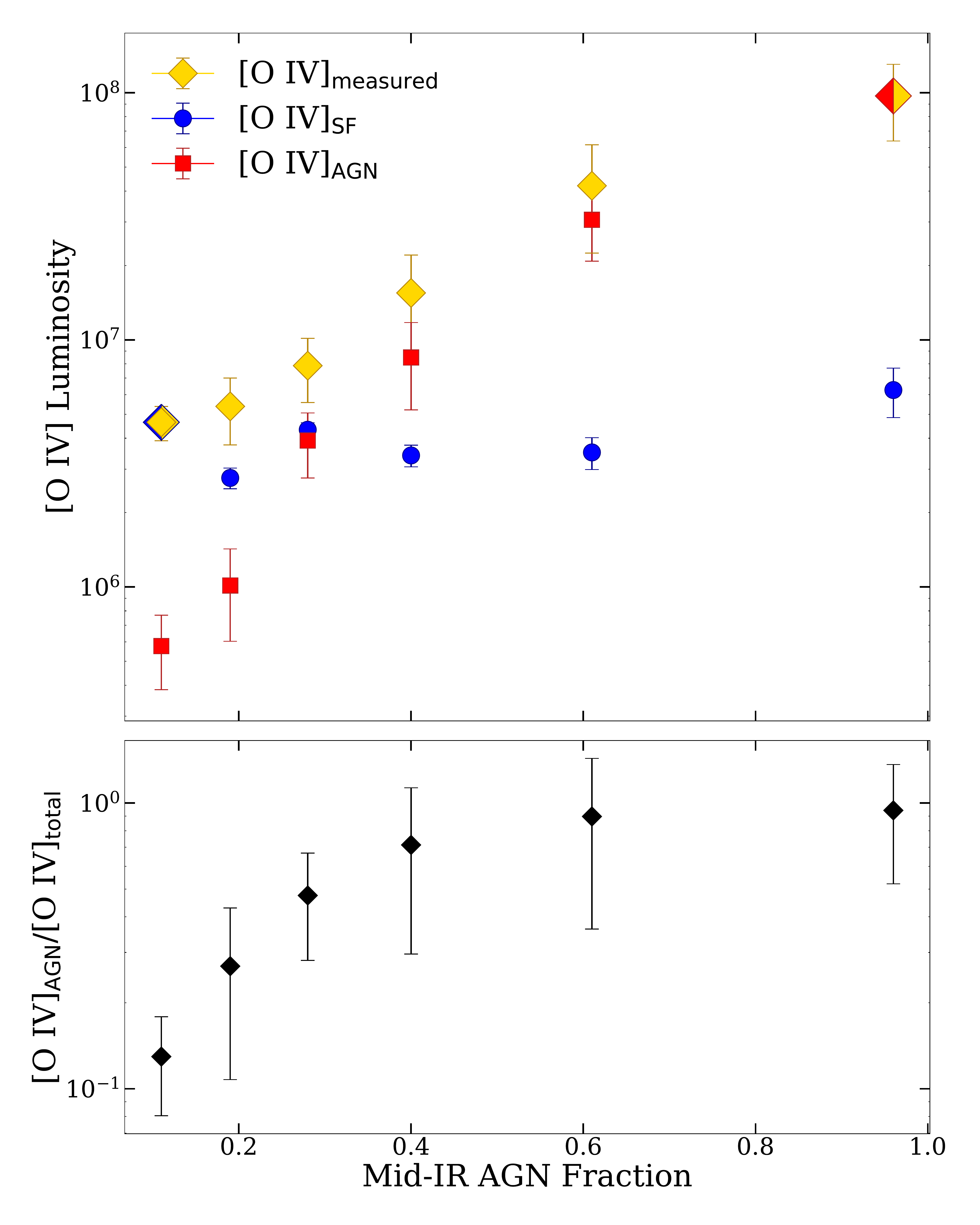}
    \caption{({\it top panel}) Decomposition of [O~IV]$_{\mathrm{measured}}$ (gold diamonds) into [O~IV]$_{\mathrm{SF}}$ (blue circles) and [O~IV]$_{\mathrm{AGN}}$ (red squares) as a function of AGN fraction. The contribution to the total [O~IV] luminosity from star formation is close to constant across the sample, while the AGN contribution ({\it bottom panel}) increases rapidly with AGN fraction, and dominates above an AGN fraction of $\gtrsim0.3$.
    }
    \label{fig:oivcontributions}
\end{figure}

[O~IV]$_{\mathrm{measured}}$, [O~IV]$_{\mathrm{SF}}$, and [O~IV]$_{\mathrm{AGN}}$ are plotted as a function of AGN fraction in Figure \ref{fig:oivcontributions}. The contribution from star formation to the [O~IV] luminosity is constant across all AGN fractions and dominates for AGN fractions $\lesssim0.3$, while the AGN contribution increases with AGN fraction and dominates for AGN fractions $\gtrsim0.3$.  

\section{Discussion} \label{sec:discussion}

\subsection{Predicting [O IV] for Individual Sources}

\begin{figure}
    \centering
    \includegraphics[width=8.33cm]{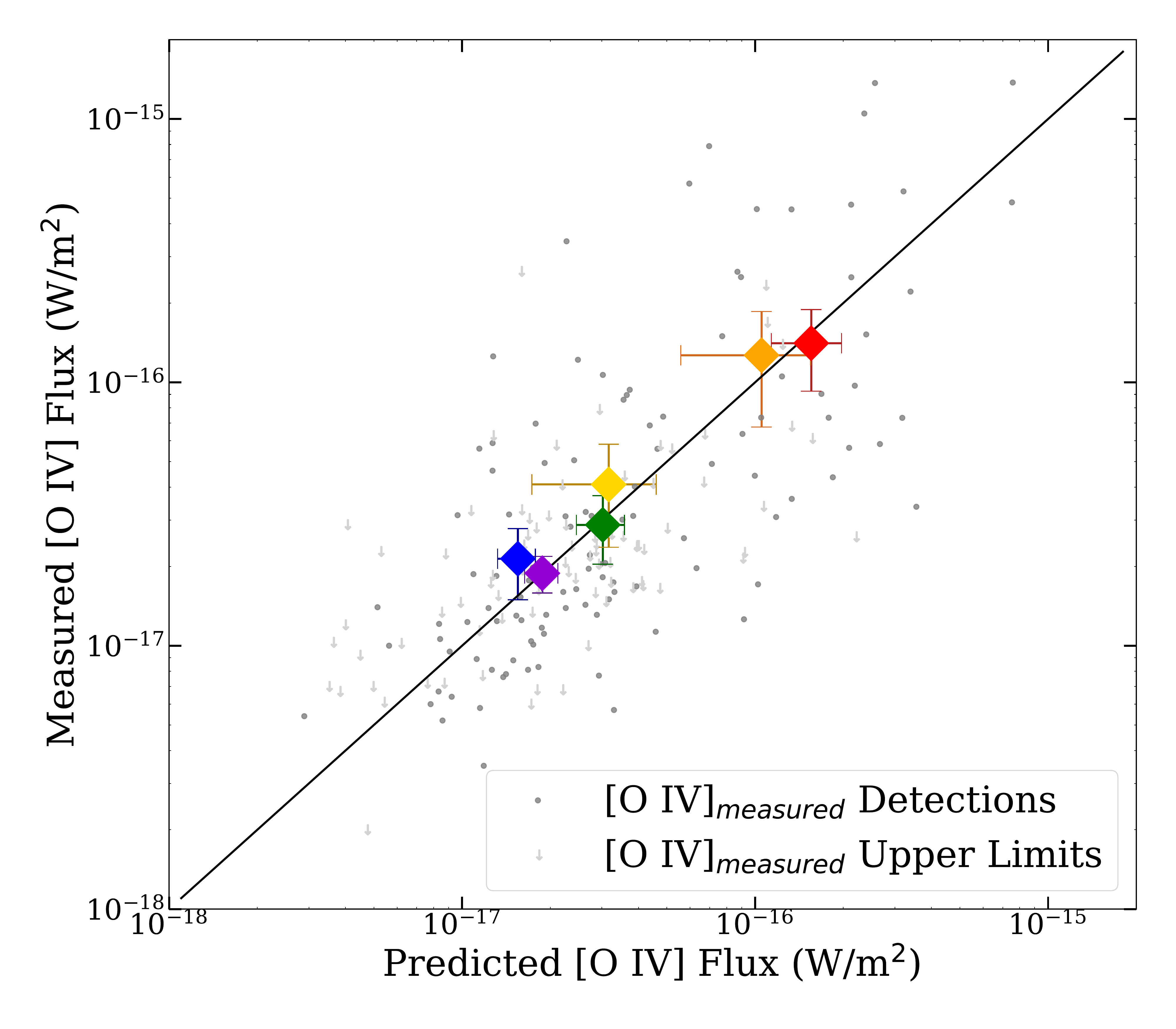}
    \caption{The predicted vs. measured [O IV] fluxes for the individual sources in the GOALS sample, and the bootstrapped flux for our six AGN fraction bins (colored diamonds). The black line is the 1-1 line, and is indistinguishable from a best fit to our six stacked spectra.}
    \label{fig:inamipredictedoiv}
\end{figure}

Using Equation \ref{eq:oivpred}, we can predict the [O IV] flux for individual GOALS sources using [Ne~II] measurements from \cite{Inami2013} and AGN fractions from \cite{DiazSantos2017}. The individual measured and predicted [O~IV] fluxes are shown alongside our stack measurements in Figure \ref{fig:inamipredictedoiv}. While there is scatter, our empirical formula, which depends only on [Ne~II] and AGN fraction, provides a good estimate of the [O~IV] flux for the full range of GOALS galaxies. This empirical formula can be useful for estimating [O~IV] fluxes for future studies of LIRGs when only the [Ne~II] and mid-IR AGN fraction are known, or for estimating the contribution of the AGN to [O~IV] using [O~IV]$_{\mathrm{AGN}}$ = [O~IV]$_{\mathrm{measured}}$ - [O~IV]$_{\mathrm{SF}}$.

\subsection{Black Hole Accretion Rates}

In samples of AGN, the [Ne V] and [O IV] lines have been shown to be good tracers of the BHAR. \cite{Dasyra2008} used a sample of well-studied nearby AGN with black hole masses measured from reverberation-mapping to show that [Ne V] and [O IV] were correlated with the AGN bolometric luminosity. Both \cite{Melendez2008} and \cite{Rigby2009} independently found the [O IV] luminosity was well-correlated with the X-ray luminosity in samples of nearby Seyferts.
\cite{Gruppioni2016} examine a sample of 12-$\mu$m selected AGN and find a linear relation between the [O~IV] and AGN luminosities:

\begin{equation}
\label{eq:LAGN}
    \log({L_{AGN}}) = (1.69 \pm 0.02)\log({L_{[O IV]}}) - (2.02 \pm 0.27)
\end{equation}

\begin{figure}
    \centering
    \includegraphics[width=8.33cm]{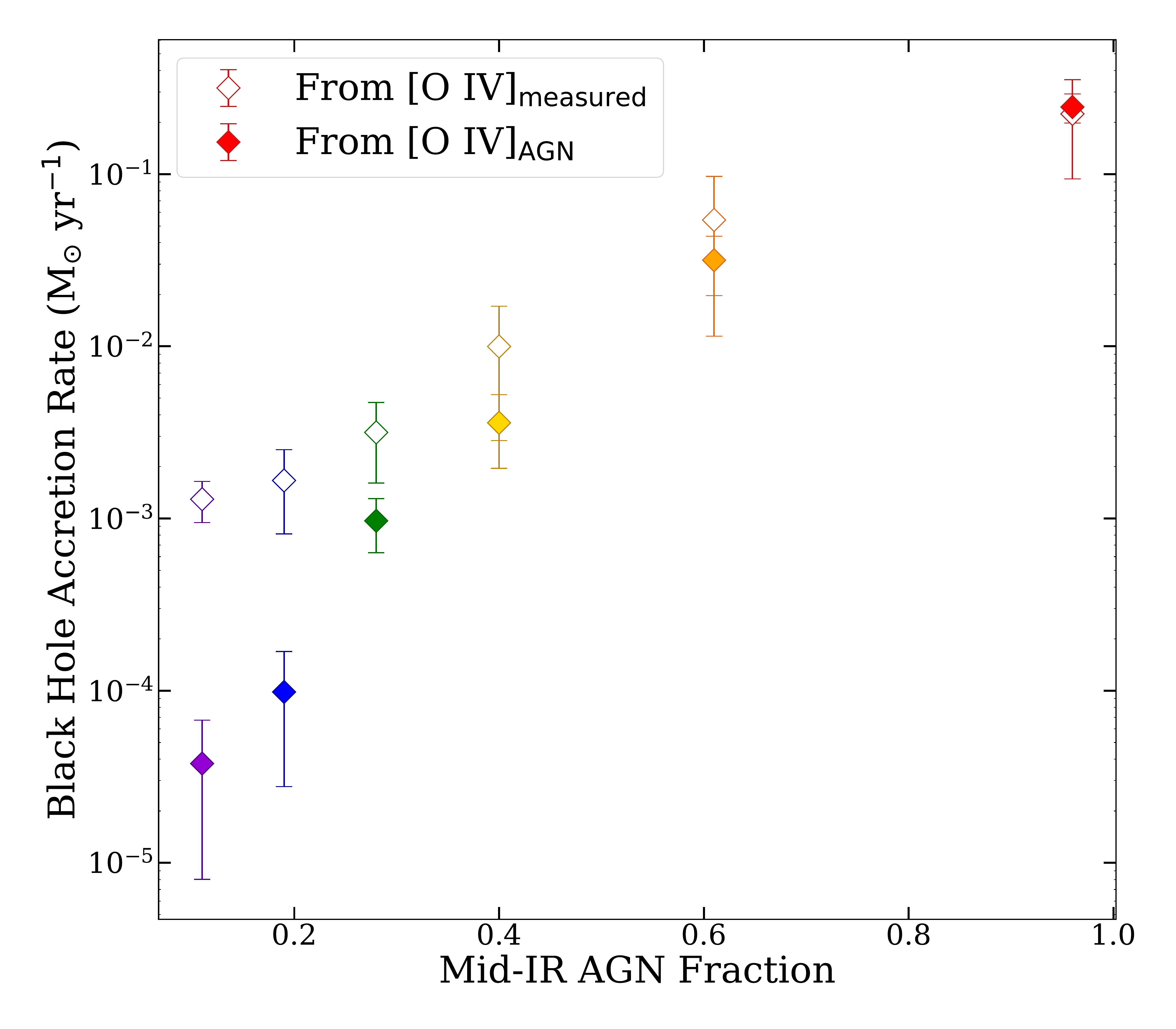}
    \caption{The black hole accretion rate as a function of mid-IR AGN fraction calculated from our total measured [O~IV] (open symbols) and the estimated [O~IV]$_{\mathrm{AGN}}$ (filled symbols) derived as [O~IV]$_{\mathrm{AGN}}$ = [O~IV]$_{\mathrm{measured}}$ - [O~IV]$_{\mathrm{SF}}$. The increase in BHAR with AGN fraction is stronger than that for [O~IV]$_{\mathrm{AGN}}$ (Figure \ref{fig:oivcontributions}) due to the steep relationship between [O~IV] and $L_{AGN}$ (see Equation \ref{eq:LAGN}).
    }
    \label{fig:bhar}
\end{figure}

Given our decomposition of [O~IV] into contributions from AGN and star formation (Section \ref{sec:decompose}), we use Equation \ref{eq:LAGN} to calculate the average BHAR as a function of AGN fraction, propagating through the standard error on the measured [O~IV] and [Ne~II] luminosities to determine the uncertainty on [O~IV]$_{\mathrm{AGN}}$ and therefore on the BHAR.
We apply our relation to our [O~IV]$_{\rm{AGN}}$, as well as to our total measured [O~IV] to quantify the difference in the absence of this correction. We convert the resulting $L_{AGN}$ to BHARs using $L_{AGN} \;(L_{\odot}) = \epsilon \dot{M}_{BH} (\mathrm{M}_{\odot}/yr) c^2$, adopting $\epsilon = 0.1$, the typical mass-energy conversion efficiency in the local Universe \citep{Marconi2004}. The BHARs derived from [O~IV]$_{\rm{AGN}}$, reported in Table \ref{tab:bharsfr} and plotted in Figure \ref{fig:bhar}, range from roughly $10^{-4}$ to $10^{-1}$ and show a steady increase as a function of AGN fraction. The BHARs derived from the total measured [O~IV], in contrast, are overestimated at low AGN fractions: by factors ranging from 30 in Bin 1 to 3 in Bin 4 due to contamination of [O~IV] by star formation. \cite{DiamondStanic2012} and \cite{PereiraSantaella2010} caution against using [O~IV] as a tracer of BHAR when the IR luminosity from star formation exceeds that from the AGN by an order of magnitude. We find that star formation contributes at least half of the [O~IV] luminosity below an AGN fraction of $\sim0.3$. This is consistent with the prescriptions of this previous work, but this analysis allows careful decomposition to be performed in this regime to overcome this contamination. Specifically, careful attention should be paid to composite galaxies, defined for example as having mid-IR AGN fractions $0.2 - 0.8$ \citep{Kirkpatrick2015}, as the contribution from star formation to the [O~IV] luminosity predicted by Equation \ref{eq:oivpred} falls from 70\% to 10\% across this range.

We propose using Equation \ref{eq:oivwnev} to calculate the contamination from star formation to [O~IV] in order to use this line as a tracer of BHAR down to low AGN fractions. In particular, by combining Equations \ref{eq:oivwnev} and \ref{eq:LAGN} we derive an expression for the corrected BHAR:

\begin{multline}
\label{eq:BHAR}
    \dot{M}_{BH} (M_{\odot}/yr) = 6.44 \times 10^{-15} \; (L_{\mathrm{[O IV]}meas} \; \\ - aL_{\mathrm{[Ne II]}} \; \, )^{1.69}
\end{multline}

where the [O~IV] and [Ne~II] luminosities are in units of $L_{\odot}$, $aL_{\mathrm{[Ne\,II]}} = \mathrm{[O\;IV]}_{\mathrm{SF}}$, and $a$ is the value derived in Section \ref{sec:decompose} ($2.865 \times 10^{-2}$ for unscaled [Ne II] fluxes).

\cite{Esquej2014} calculated black hole accretion rates for a sample of nearby Seyfert galaxies, finding black hole accretion rates ranging from $5 \times 10^{-6}$ to 0.5 M$_{\odot}$/yr, very similar to the BHARs we find in the GOALS sample.
\cite{Yang2017} determine the average BHAR in bins of SFR for $\sim18,000$ galaxies in the CANDELS/GOODS-South field from $0.5 \leq z < 2.0$ using X-ray observations. The average \cite{Yang2017} BHAR for the bins with star formation rates similar to our sample is $\sim 5 \times 10^{-3}$, in general agreement with our values.

\subsection{Star Formation Rates}

\begin{figure*}
    \centering
    \includegraphics[width=16.5cm]{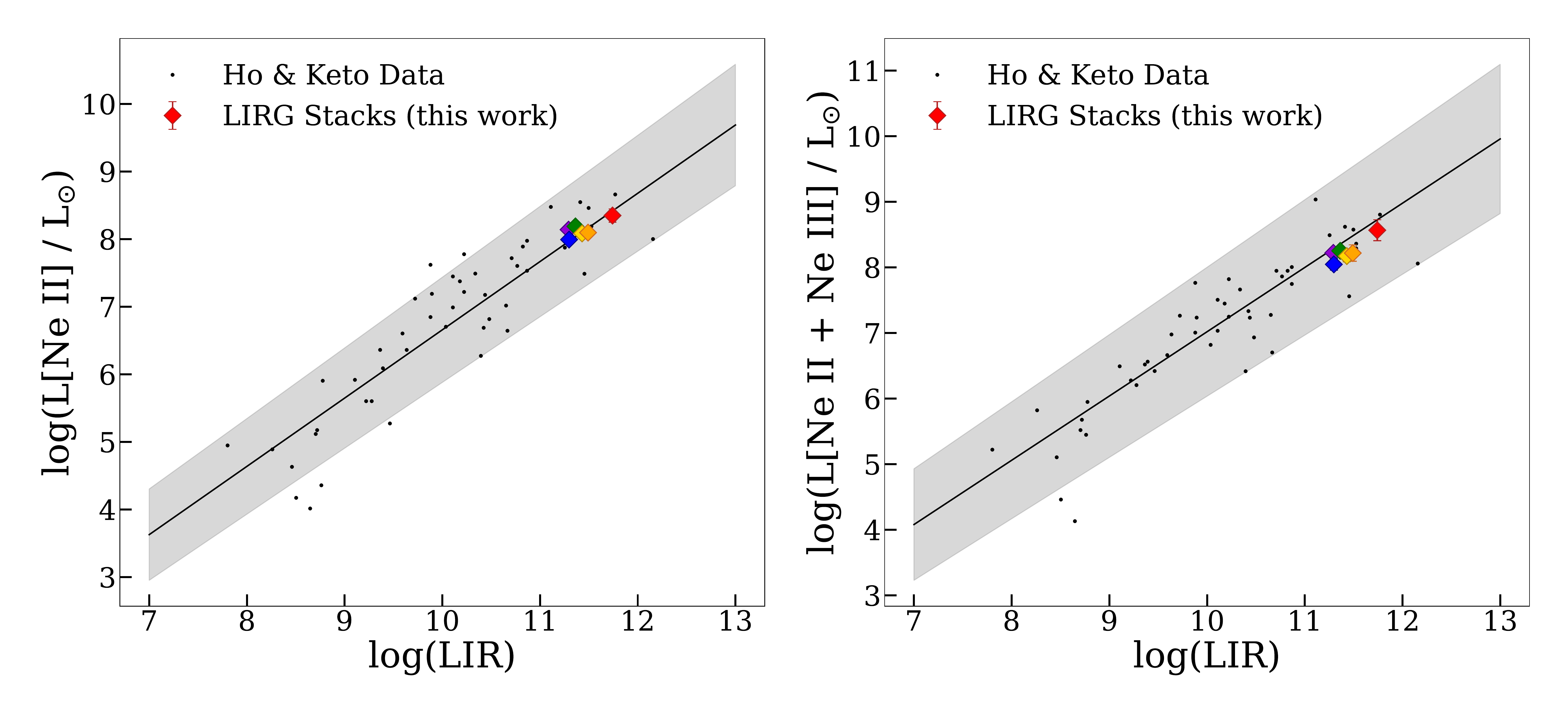}
    \caption{The [Ne~II] \textit{(left)} and [Ne~II] + [Ne~III] \textit{(right)} luminosities as a function of $L_{\mathrm{IR}}$, compared to the data and relations from \cite{Ho2007}. Our six AGN fraction stacks lie well within the scatter for both [Ne~II] and [Ne~II] + [Ne~III].
    }
    \label{fig:hoketocomparison}
\end{figure*}

\cite{Ho2007} show that the luminosities of [Ne~II] and [Ne~II] + [Ne~III] in star-forming galaxies correlate strongly with $L_{\rm{IR}}$ over several orders of magnitude. In Figure \ref{fig:hoketocomparison}, we plot the line luminosities for our six AGN fraction bins as compared to the \cite{Ho2007} relations for [Ne~II] (left panel) and [Ne~II] + [Ne~III] (right panel). Our data is well within the scatter in the \cite{Ho2007} relations, though we fall systematically slightly below the [Ne~II] + [Ne~III] fit relation. \cite{Farrah2007} find the [Ne~II] + [Ne~III] luminosity for a subset of local ULIRGs is systematically offset from the \cite{Ho2007} relation for normal star-forming galaxies, and attribute the difference to higher dust extinction in their sample of ULIRGs. The stacked measurements from our sample of both LIRGs and ULIRGs includes more low-extinction sources, and we find they are consistent with the normal star forming galaxies from \cite{Ho2007}.

\cite{Ho2007} also present a relation between the luminosity of [Ne~II] + [Ne~III] and the star formation rate \citep[Equation 13,][]{Ho2007}. We use this relation to calculate the SFRs for each AGN fraction bin assuming 75\% of a typical galaxy's neon forbidden line emission arises from singly ionized [Ne~II] and 10\% from doubly ionized [Ne~III] \citep[values from][]{Ho2007,Farrah2007}. We compare the \cite{Ho2007}, [Ne~II] + [Ne~III]-derived SFRs to SFRs calculated from $L_{\mathrm{IR,SF}}$ \citep[using the relation from][]{Murphy2011}, and find that they are consistent. 

\begin{deluxetable}{ccc}
\tablenum{4}
\tablecaption{Star formation rates and black hole accretion rates for our six AGN fraction bins. \label{tab:bharsfr}}
\tablewidth{0pt}
\tablehead{
\colhead{Bin} & \colhead{SFR (from L$_{IR,SF}$)} & \colhead{BHAR (from [O~IV]$_{AGN}$,} \\
\colhead{} & \colhead{(M$_{\odot}$/yr)} & \colhead{Eq. \ref{eq:BHAR}) (M$_{\odot}$/yr)}
}
\decimalcolnumbers
\startdata
Bin 1 & $26.3 \pm 4.8$ & $(3.8 \pm 3.0) \times 10^{-5}$ \\
Bin 2 & $25.9 \pm 2.8$ & $(9.8 \pm 7.0) \times 10^{-5}$ \\
Bin 3 & $31.8 \pm 3.4$ & $(9.7 \pm 3.4) \times 10^{-4}$ \\
Bin 4 & $33.9 \pm 5.9$ & $(3.6 \pm 1.6) \times 10^{-3}$ \\
Bin 5 & $36.4 \pm 10.0$ & $(3.2 \pm 1.2) \times 10^{-2}$ \\
Bin 6 & $40.4 \pm 8.8$ & $(2.5 \pm 0.4) \times 10^{-1}$ \\
\enddata
\end{deluxetable}

\subsection{Black Hole Accretion Rate vs. Star Formation Rate and Comparison to Literature}

Because we use the luminosity of [Ne~II] to determine [O IV]$_{\mathrm{AGN}}$ and therefore our BHARs, we use the SFR from $L_{\mathrm{IR,SF}}$ in order to keep the quantities independent: these star formation rates are listed in Table \ref{tab:bharsfr}, where the uncertainty reflects both the standard error on the median AGN fraction and L$_{IR}$ of each bin.

\begin{figure}
    \centering
    \includegraphics[width=8.33cm]{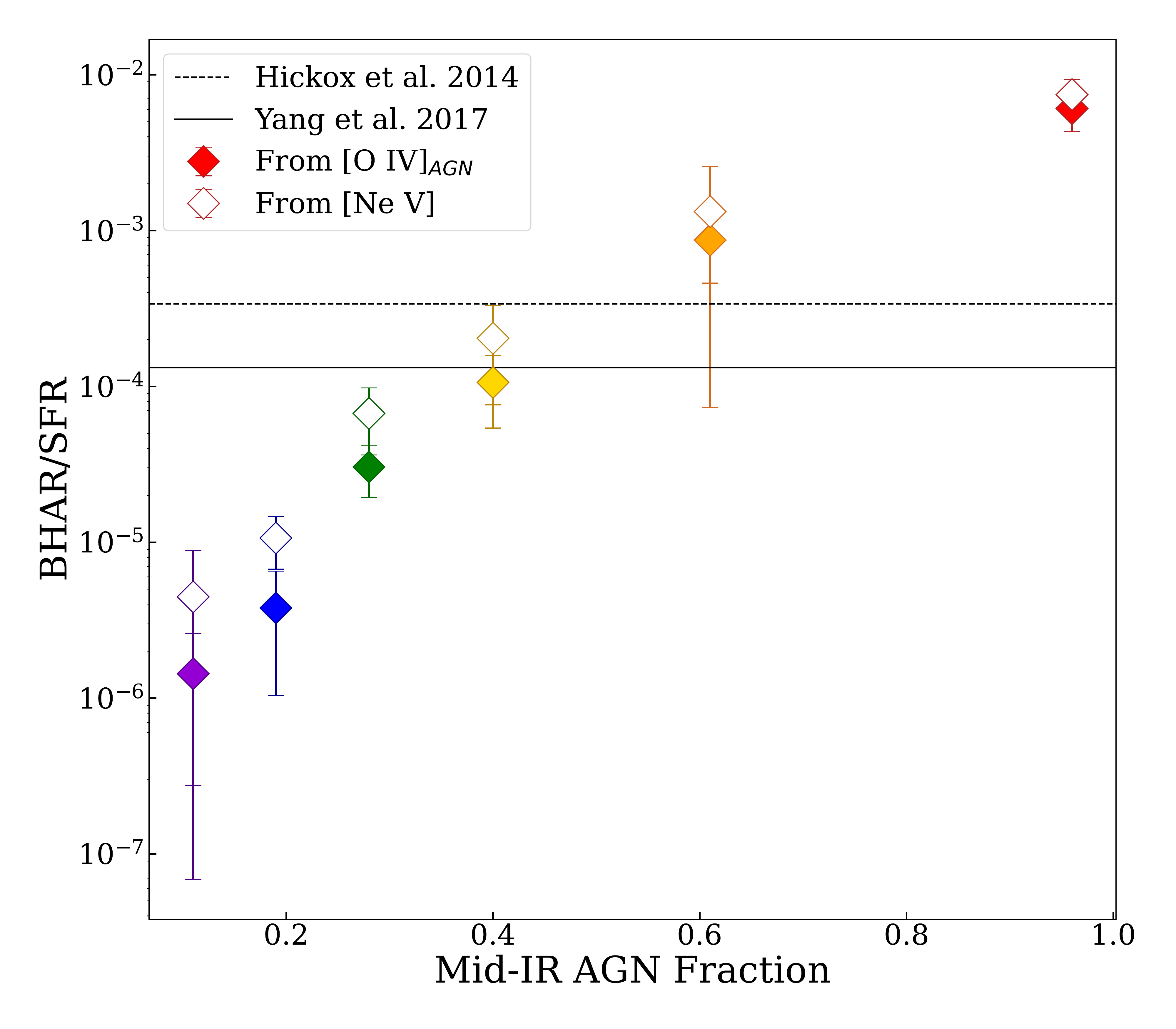}
    \caption{The ratio of BHAR/SFR versus the AGN fraction for our six bins of AGN fraction. We derive the BHAR from OIV$_{\mathrm{AGN}}$ and the SFR from L$_{\mathrm{IR,SF}}$. We include two median BHAR/SFR values from the literature: the \cite{Yang2017} value (solid line), from a sample containing a mixture of star-forming galaxies and AGN, and the \cite{Hickox2014} value, derived from a model where each modeled galaxy has an AGN component. Both literature results are consistent with our intermediate bins. We also plot the BHAR derived from our predicted [Ne~V] (Equation \ref{eq:ratio}) and \cite{Gruppioni2016} (open circles), which yield similar values.
    }
    \label{fig:BHARSFRAGN}
\end{figure}

In Figure \ref{fig:BHARSFRAGN}, we plot the ratio of the BHAR (from Equation \ref{eq:BHAR}) and SFR (from $L_{\mathrm{IR,SF}}$) as a function of the mid-IR AGN fraction. We note that this ratio increases by 3--4 orders of magnitude over the full range of AGN fractions, suggesting a wide range of relative growth between the supermassive black hole and stars within the GOALS sample. 

We overplot two estimates of BHAR/SFR from the literature. In the \cite{Hickox2014} model (dotted line), a time-averaged (100 Myr) BHAR/SFR of $3.33 \times 10^{-4}$ is adopted to demonstrate that a correlated long-term BHAR and SFR reproduce the weak correlations observed between SFR and $L_{\mathrm{AGN}}$. 
\cite{Yang2017} find a slightly lower median BHAR/SFR of $1.41 \times 10^{-4}$ (with average ratios in bins of M$_*$ ranging from $\sim10^{-5}$ to $10^{-3}$), using BHARs derived from X-ray observations and SFRs from SED fitting. Both of these BHAR/SFR ratios are consistent with our intermediate bins, or the average over all AGN fraction bins. Both \cite{Hickox2014} and \cite{Yang2017} demonstrate a large dynamic range of BHAR/SFR ratios; our study confirms this wide range continues down to the local LIRG regime, and correlates strongly with the mid-IR AGN fraction.

We also apply the \cite{Gruppioni2016} [Ne~V]--$L_{AGN}$ relation to our predicted [Ne~V] (Equation \ref{eq:ratio}) for comparison. These two independent calculations of BHAR return results consistent within a factor of 2 for all AGN fraction bins. We note that if we assume the [Ne V] luminosity is instead constant below a mid-IR AGN fraction of $\sim 0.3$, the ratio BHAR/SFR still increases by two orders of magnitude; however, the high level of agreement between [O~IV]$_{\mathrm{predicted}}$ and [O IV]$_{\mathrm{measured}}$ in Bins 1 and 2 in the rightmost panel of Figure \ref{fig:fittingoiv} support the extrapolation of our empirical fit to low [Ne V]/[Ne II] ratios in the domain of the upper limits at low mid-IR AGN fraction.

\section{Summary and Future Applications}
\label{sec:summary}
We perform a spectral stacking analysis as a function of mid-IR AGN fraction using 203 \textit{Spitzer}/IRS spectra of the GOALS sample in order to explore the co-evolution of the mid-infrared atomic emission lines $12.8\,\mu$m [Ne~II], $14.3\,\mu$m [Ne~V], $15.5\,\mu$m [Ne~III], and $25.9\,\mu$m [O~IV].
\begin{itemize}[leftmargin=*]
    \item We detect all lines in all bins except for [Ne~V] in our two lowest-AGN fraction stacked spectra, where we derive 3$\sigma$ upper limits.
    \item We compare line ratios to results from the literature for other samples of galaxies, confirming that our low-AGN fraction stacked samples are consistent with H II galaxies and our high-AGN fraction bins most closely resemble LINER or Seyfert 2 galaxies.
    \item We derive a relationship between the [Ne~II] luminosity, the AGN fraction, and the [O~IV] luminosity, allowing us to determine the relative contributions to [O~IV] from star formation and AGN. This relation can be used to correct for the contamination of [O~IV] due to star formation given the AGN fraction. 
    \item We apply this empirical relation to predict the [O~IV] luminosity of individual galaxies in the GOALS samples and find excellent agreement with the measurements. 
    \item Using our [O~IV] decomposition, we calculate black hole accretion rates for stacked spectra from [O~IV]$_{\mathrm{AGN}}$, and quantify the evolution of the BHAR/SFR ratio as a function of AGN fraction. This ratio spans over three orders of magnitude across the GOALS sample.
\end{itemize}

We have used the technique of a spectral stacking analysis to detect faint star formation signals in AGN-dominated sources and faint AGN signals in star formation-dominated sources. We intend to apply this technique to study additional spectral lines in the low and high resolution spectra including the PAH lines, molecular hydrogen (H$_2$) rotational lines and fainter atomic lines such as [Cl~II] and [Fe~II]. The differences in the distribution of merger stages between our bins motivate extending our stacking analysis to examine the evolution of mid-infrared spectral features as a function of merger stage. 

The {\it James Webb Space Telescope} ({\it JWST}) will provide renewed access to the mid-infrared spectral regime at unprecedented spectral resolution and sensitivity. Our stacking analysis provides average mid-IR line fluxes for IR-selected galaxies as a function of AGN fraction which are helpful for planning {\it JWST} observations. Our empirical formula can be useful for estimating [O~IV] fluxes for future studies of LIRGs when only the [Ne~II] and mid-IR AGN fraction are known. In addition to the key star formation and AGN line measurements presented in this paper, we make our stacked spectral publicly available\footnote{\url{https://people.astro.umass.edu/~pope/Stone2022/}}. With the {\it JWST}/MIRI IFU, the mid-IR lines can be spatially resolved in individual nearby galaxies, with [Ne~V] and [Ne~II] both observable in galaxies out to $z\lesssim1$. Furthermore, the spectral resolution of {\it JWST} can be exploited to study the kinematics of the gas \citep{Inami2013}. The physical extent and velocity of the star formation and AGN line emission will provide further diagnostics on the coevolution of these processes in galaxies. In the longer term, a new, cold far-IR space telescope will be required to probe these powerful diagnostics in galaxies over all cosmic time to map the relative buildup of stars and supermassive black holes.

\acknowledgments
We thank the referee for constructive comments which improved this work. We thank Krista Gile for advice regarding statistics and error analysis.
MS is grateful for support from the Massachusetts Space Grant consortium, the Commonwealth Honors College, and the Five College Astronomy Department.
This research has made use of the NASA/IPAC Infrared Science Archive, which is funded by the National Aeronautics and Space Administration and operated by the California Institute of Technology. We thank the IRSA staff for help with the GOALS IRS data products.

\bibliography{refs}{}
\bibliographystyle{aasjournal}

\end{document}